\documentclass[twocolumn,resetfootnote]{aastex701}
\shorttitle{NPE for Inferring Weak Lensing Shear}
\shortauthors{White et al.}
\usepackage{graphicx}
\usepackage{amsmath}
\usepackage{subcaption}
\usepackage{tikz}
\usepackage{booktabs}
\usepackage{multirow}
\usetikzlibrary{positioning,shapes.geometric,bending,fit}
\usepackage{caption}
\usepackage[nameinlink,capitalize,noabbrev]{cleveref}
\creflabelformat{equation}{#2#1#3}
\crefname{appendix}{Appendix}{Appendices}
\Crefname{appendix}{Appendix}{Appendices}
\usepackage{etoolbox}
\makeatletter
\apptocmd\@sect{%
  \cref@constructprefix{#1}{\cref@result}%
  \@ifundefined{cref@#1@alias}%
    {\def\@tempa{#1}}%
    {\def\@tempa{\csname cref@#1@alias\endcsname}}%
  \protected@edef\cref@currentlabel{%
    [\@tempa][\arabic{#1}][\cref@result]%
    \csname p@#1\endcsname\csname the#1\endcsname}%
}{}{\fail}
\makeatother

\usepackage{todonotes}
\setuptodonotes{inline}

\newcommand{\padvert}{\thinspace \vert \thinspace}
\definecolor{mediumteal}{RGB}{9, 123, 125}
\definecolor{pinkgray}{RGB}{122, 104, 119}

\begin{document}

\title[NPE for Inferring Weak Lensing Shear]{Neural Posterior Estimation for Inferring Weak Lensing Shear}

\author[orcid=0000-0001-5535-0452]{Tim White}
\affiliation{Department of Statistics, University of Michigan}
\email{twhit@umich.edu}

\author[orcid=0009-0009-9783-0839]{Dingrui Tao} 
\affiliation{Department of Statistics, University of Michigan}
\email{taodingr@umich.edu}

\author[orcid=0000-0001-8868-0810]{Camille Avestruz}
\affiliation{Department of Physics, University of Michigan}
\email{cavestru@umich.edu}

\author[orcid=0000-0002-1472-5235]{Jeffrey Regier}
\affiliation{Department of Statistics, University of Michigan}
\email{regier@umich.edu}

\collaboration{all}{The LSST Dark Energy Science Collaboration}

\begin{abstract}
The prevailing approach to inferring weak gravitational lensing shear from images involves detecting galaxies, estimating their ellipticities, and calibrating these estimates to correct for image noise, selection bias, and model misspecification. Characterizing the statistical model and assumptions underlying this pipeline is challenging, which makes it difficult to propagate uncertainty through its various stages. As an alternative, we propose to infer shear using neural posterior estimation (NPE), a type of simulation-based inference. We train a deep neural network to map a simulated multiband image to a variational distribution over the underlying shear field, thereby folding galaxy detection, deblending, measurement, and calibration into a single implicit inference step. Once trained, the network accounts for all features present in the simulated images, including potential sources of bias. In experiments on simulated constant-shear images with increasingly complex observational effects, NPE produces accurate and well-calibrated posterior approximations for both shear components in the presence of blended galaxies, spatially varying point spread functions, stars, and detector artifacts. These results demonstrate that NPE can be a viable shear estimation method in settings where all anticipated features and artifacts can be simulated, a requirement that will become increasingly feasible as simulation fidelity improves in the coming decades.
\end{abstract}



\section{Introduction}

Weak gravitational lensing refers to the spatially correlated distortion of the apparent shapes of distant galaxies by intervening matter. This matter curves spacetime, deflecting photon trajectories and inducing a subtle but measurable transformation of galaxy images. The resulting distortion field, typically parametrized by shear (i.e., anisotropic stretching) and convergence (i.e., isotropic scaling), encodes information about the projected intervening matter distribution, making weak lensing a powerful probe of the Universe's large-scale structure \citep{bartelmann2001weak, kilbinger2015cosmology, dodelson2017gravitational, mandelbaum2018weak, prat2025weak}.

Stage IV surveys such as the Vera C. Rubin Observatory's Legacy Survey of Space and Time (LSST; \citealt{ivezic2019lsst}), the Euclid Space Telescope \citep{laureijs2011euclid}, and the Nancy Grace Roman Space Telescope \citep{spergel2015wfirst, akeson2019roman, dore2019wfirst} will survey billions of galaxies. Constraining cosmological parameters based on these surveys will require precise shear measurements.

Precisely measuring shear is challenging due to numerous observational effects which, unless modeled or corrected for, induce systematic biases in shear estimates \citep{jain2006psf,mandelbaum2015great}. For example, point spread functions (PSFs) convolve galaxy images, diluting and distorting the shear signal \citep{zhang2022impact, zhang2023impact, liaudat2023psf}. The PSF typically varies across the telescope's focal plane, which makes correcting for the resulting convolution challenging. Image noise further biases the shear signal in each exposure \citep{melchior2012noise, gurvich2016impact}, and correcting for this effect is challenging because exposures are typically combined, or ``coadded'' \citep{mandelbaum2023psfs, armstrong2024little}. Many imaged galaxies visually overlap, or ``blend,'' complicating both detection and shape measurement \citep{melchior2021challenge, sanchez2021effects, zhang2025blending}. Sample selection based on observed properties, such as signal-to-noise ratio or size, can introduce shear-dependent biases \citep{mandelbaum2005systematic,huff2014seeing, jarvis2016science,fenech2017calibration, sheldon2020mitigating}. Artifacts such as saturated stars with bleed trails, cosmic rays, and bad charge-coupled device (CCD) columns must be identified before images can be processed \citep{massey2013origins,mandelbaum2015instrumental}.

The conventional approach to shear estimation involves a multistage pipeline. First, galaxies are detected and their ellipticities are measured. Then, shear is estimated by computing an intricate weighted average of these ellipticities, and the shear estimates are calibrated to correct for biases. Metadetection performs this calibration empirically by artificially shearing images and measuring the estimator's response \citep{sheldon2017practical, sheldon2020mitigating, sheldon2023metadetection, yamamoto2025dark}. The Analytical Calibration method (or AnaCal) derives this calibration analytically \citep{li2023analytical, li2024differentiable, li2025analytical}.

Alternative approaches target the posterior distribution of shear conditional on the image pixels. The Bayesian Fourier Domain (BFD) paradigm uses a Bayesian framework with grid-based likelihood evaluation \citep{bernstein2014bayesian, bernstein2016accurate}. Hierarchical Bayesian methods use MCMC to sample the joint posterior over shear and galaxy properties \citep{schneider2015hierarchical, congedo2024euclid, mendoza2026differentiable}. While these probabilistic approaches provide richer uncertainty quantification than methods that produce only point estimates, they are not amortized: BFD evaluates the likelihood on a grid, and hierarchical methods rely on MCMC. Thus, the cost of inference scales with the number of galaxies analyzed, a significant computational challenge for surveys that will image billions of galaxies.

For Stage IV surveys, observational effects are expected to be one of the primary obstacles to precise shear measurement, along with astrophysical systematics such as intrinsic alignments \citep{mandelbaum2018weak}. All of the methods described above account for observational effects explicitly, either by absorbing them into a calibration step or incorporating them into a likelihood function. The analytical and computational cost of doing so will grow as the precision required of shear estimators increases in the coming years.

As an alternative, we propose to infer shear using neural posterior estimation (NPE), a type of amortized variational inference. This procedure involves training a neural network to map a simulated multiband image to an approximate posterior distribution over shear. With this approach, observational effects must be simulated in the training images, but they otherwise require no explicit correction. Additionally, NPE amortizes the cost of inference: estimating shear for new images requires only a forward pass of the trained network.

In this work, we evaluate NPE on constant-shear images simulated using \texttt{descwl-shear-sims}, a software package that emulates LSST observing conditions (\cref{sec:simulatedimages}). We assess the capabilities of NPE for a particular mean-field variational family and deep residual neural network architecture (\cref{sec:npe}). Under these choices, the trained network produces accurate and well-calibrated posterior approximations for both shear components in the presence of blended galaxies, spatially varying PSFs, stars (including saturated stars), and CCD artifacts (\cref{sec:traineval,sec:experiments}). While these experiments demonstrate the robustness of NPE to observational effects when the training and evaluation images come from the same distribution, they do not address the possibility of distribution shift. We discuss methods to guard against misspecification of the simulator's underlying probabilistic model, and we connect the sensitivity analysis performed in this paper to recent work that uses NPE to infer tomographic, spatially varying shear and convergence maps (\cref{sec:discussion}).

\section{Simulated images} \label{sec:simulatedimages}

We simulate multiband exposures using \texttt{descwl-shear-sims}, a software package designed to prepare shear estimation algorithms for LSST \citep{sheldon2023metadetection}. Our simulation procedure is similar to those detailed in \cite{sheldon2023metadetection} and \cite{li2025analytical}, which use the same package to evaluate Metadetection and AnaCal, respectively. Starting from a galaxy-only baseline with a fixed point spread function, LSST-like noise, and no detector artifacts (\cref{subsec:baseline}), we introduce observational effects one at a time. We first switch to a spatially varying PSF (\cref{subsec:psf}). Next, we incorporate stars (\cref{subsec:stars}), followed by cosmic rays and bad CCD columns (\cref{subsec:artifacts}). Finally, we reduce the galaxy density to one-third of its initial value. \Cref{table:simulationsettings} summarizes our five simulation settings.

Several simplifying assumptions underlie all of these settings: the image noise is independent across pixels in the coadd world coordinate system, and all galaxies in each image have the same redshift, the same shear, and a convergence of zero. We discuss the implications of these assumptions in \cref{subsec:limitations}.

\begin{table*}
    \centering
    \begin{tabular}{ccccc}
        \toprule
        & \textbf{PSF} & \textbf{Stars} & \textbf{Image artifacts} & \textbf{Galaxies per arcmin$^2$} \\
        \midrule
        \textbf{Setting 1} & Fixed    & No  & No  & 240 \\
        \textbf{Setting 2} & Variable & No  & No  & 240 \\
        \textbf{Setting 3} & Variable & Yes & No  & 240 \\
        \textbf{Setting 4} & Variable & Yes & Yes & 240 \\
        \textbf{Setting 5} & Variable & Yes & Yes & 80  \\
        \bottomrule
    \end{tabular}
    \caption{For all settings, we simulate $2048{\times}2048$-pixel images in the $riz$ bands, with galaxies generated using the \texttt{WeakLensingDeblending} package (\cref{subsec:baseline}). The fixed point spread function (PSF) uses a circular Moffat profile, and the spatially varying PSF model is based on the power-spectrum-based method of \cite{heymans2012impact} (\cref{subsec:psf}). Star densities and fluxes are sampled from the DC2 Simulated Sky Survey stellar catalog (\cref{subsec:stars}). Image artifacts include cosmic rays and bad columns (\cref{subsec:artifacts}).}
    \label{table:simulationsettings}
\end{table*}

\subsection{Baseline settings} \label{subsec:baseline}

We simulate multiband ($riz$) images of size $2048{\times}2048$ pixels with a pixel scale of 0.2 arcseconds. Each image is rendered as a single exposure on the coadd world coordinate system, with the noise amplitude in each band scaled to the projected year-10 LSST coadd depth. The noise is Gaussian and independent across pixels.

For each image, we sample the two shear components (denoted $\gamma_1$ and $\gamma_2$) independently from a Gaussian distribution with mean zero and standard deviation $0.015$. This prior distribution is based on the expected shear amplitudes across LSST patches. The sampled shears are applied to all galaxies in the image, as this is currently the only functionality provided by \texttt{descwl-shear-sims}.

We use the \texttt{WeakLensingDeblending} package \citep{sanchez2021effects} to simulate galaxies from the CatSim catalog \citep{connolly2014catsim} for all five settings. Each galaxy is modeled as a combination of Sersic disk and bulge components, and a small fraction also include an active galactic nucleus component. The morphology of each galaxy is the same across the $r$, $i$, and $z$ bands; only the apparent magnitude varies across bands. We impose an upper AB magnitude limit of 27 in the $i$ band, yielding a raw density of 240 galaxies per square arcminute, which results in substantial blending. In our final simulation setting, we reduce the raw density to 80 galaxies per square arcminute. In both cases, the effective number density is considerably lower than the raw density \citep{sanchez2021effects}.

\subsection{Point spread function} \label{subsec:psf}

In setting 1, we use a circular Moffat point spread function (PSF) with a full width at half maximum of 0.8 arcseconds and a shape parameter of 2.5, similar to \cite{sheldon2023metadetection} and \cite{li2025analytical}. In the remaining settings, we use the spatially varying PSF model proposed in \cite{heymans2012impact} and described in the appendix of \cite{sheldon2020mitigating}, which is implemented in \texttt{descwl-shear-sims}. This model uses a Moffat profile with location-dependent ellipticity, so the PSF is not circular in settings 2 through 5. The top left panel of \cref{fig:images} displays an image from setting 2.

\subsection{Stars} \label{subsec:stars}

We use the default star simulation procedure provided by the \texttt{descwl-shear-sims} package \citep{sheldon2023metadetection, li2025analytical}. Stellar densities are randomly sampled from the density map used for the LSST Dark Energy Science Collaboration's DC2 Simulated Sky Survey \citep{abolfathi2021lsst, abolfathi2021desc}. Stellar fluxes are sampled from the Milky Way star catalog used in DC2.

The top-right panel of \cref{fig:images} displays an example image for setting 3. Stars are visible throughout the image, including several saturated stars with bleed trails.

\subsection{Image artifacts} \label{subsec:artifacts}

\begin{figure*}
    \centering
    \begin{subfigure}[t]{0.4\textwidth}
        \centering
        \includegraphics[width=0.95\textwidth]{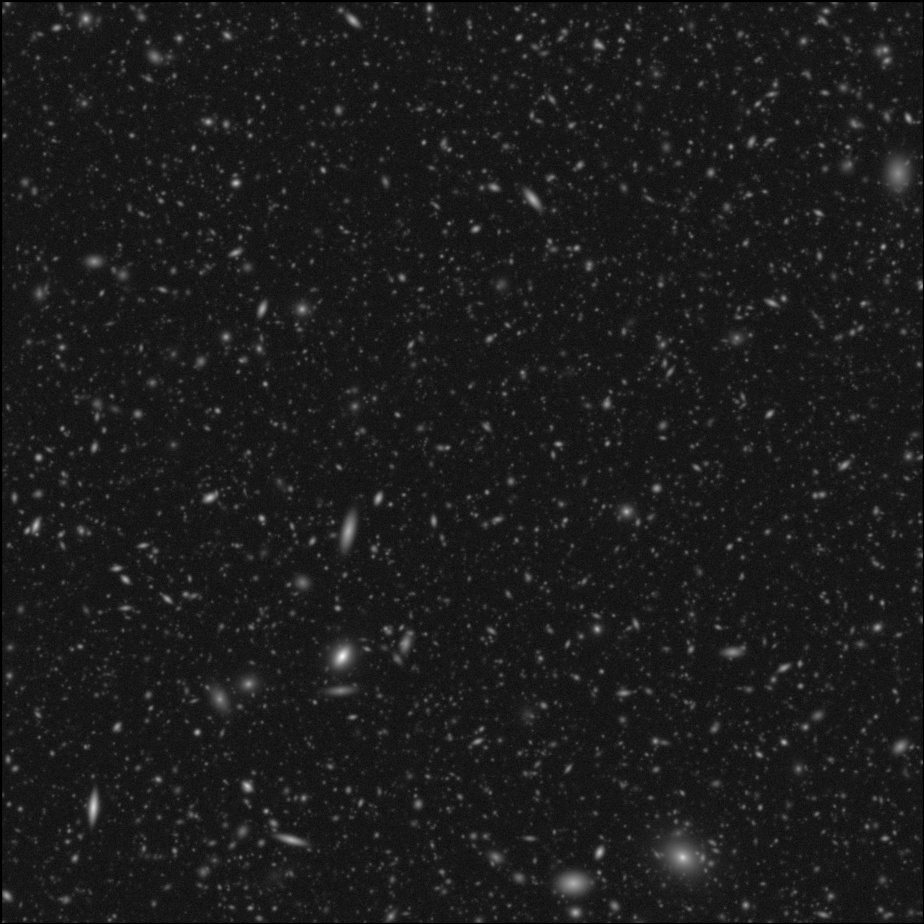}
    \end{subfigure}
    \begin{subfigure}[t]{0.4\textwidth}
        \centering
        \includegraphics[width=0.95\textwidth]{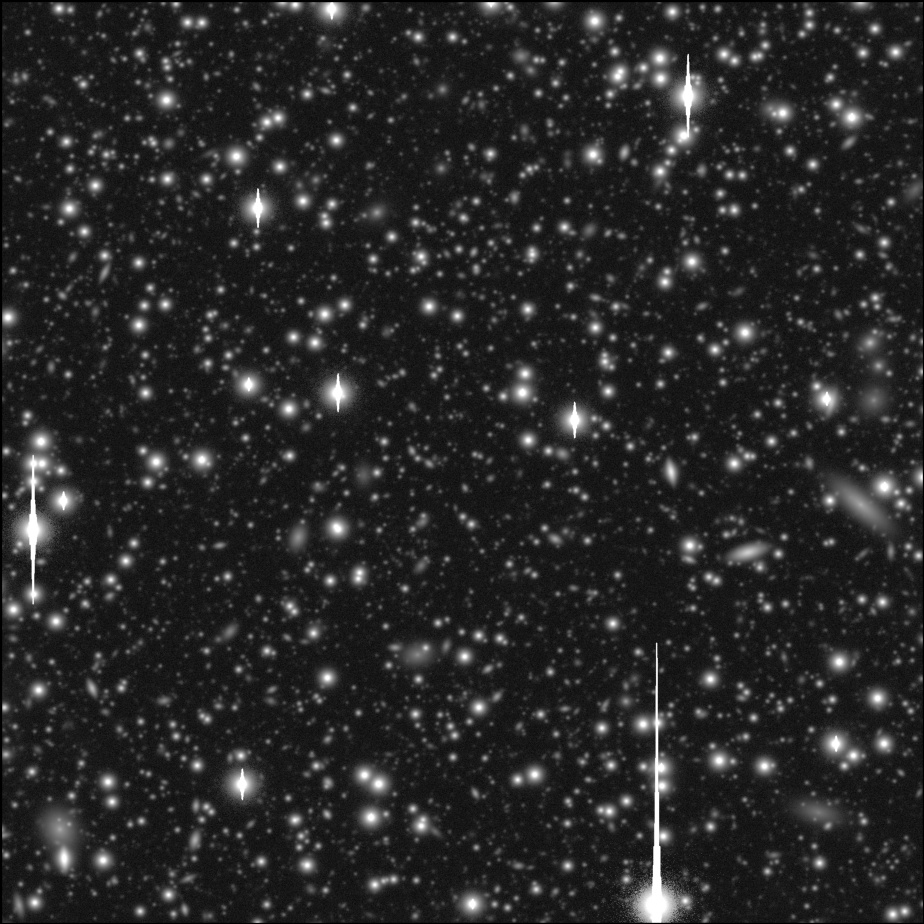}
    \end{subfigure}

    \medskip
    \medskip
    
    \begin{subfigure}[t]{0.4\textwidth}
        \centering
        \includegraphics[width=0.95\textwidth]{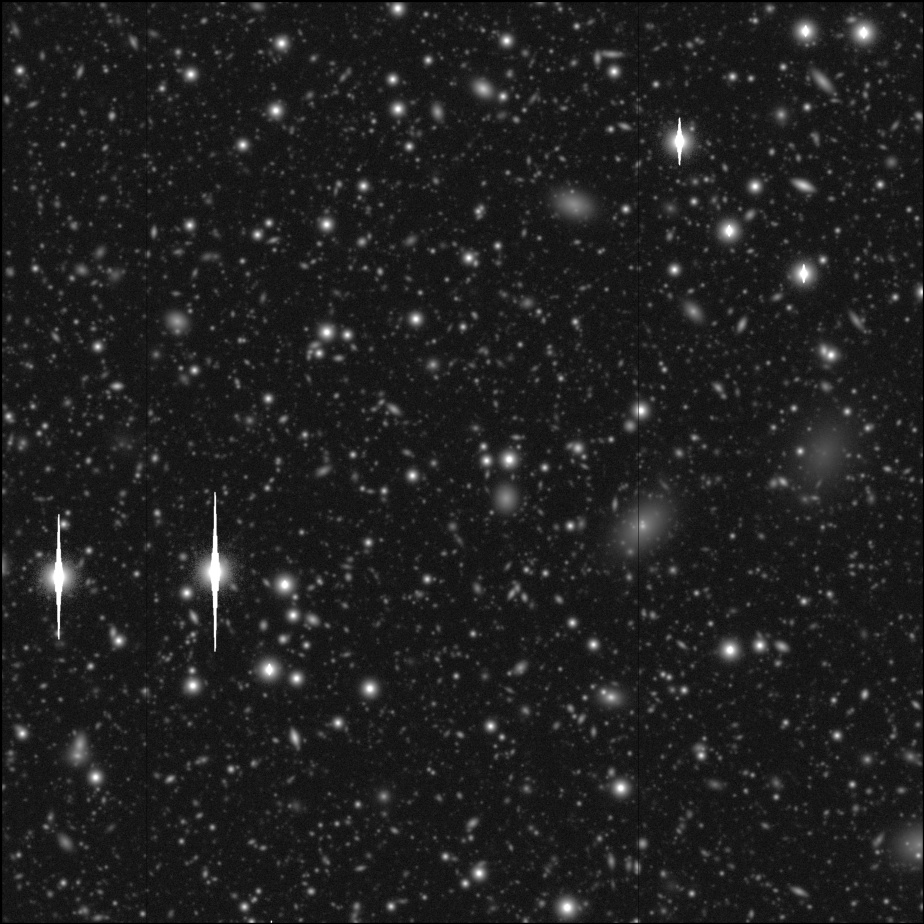}
    \end{subfigure}
    \begin{subfigure}[t]{0.4\textwidth}
        \centering
        \includegraphics[width=0.95\textwidth]{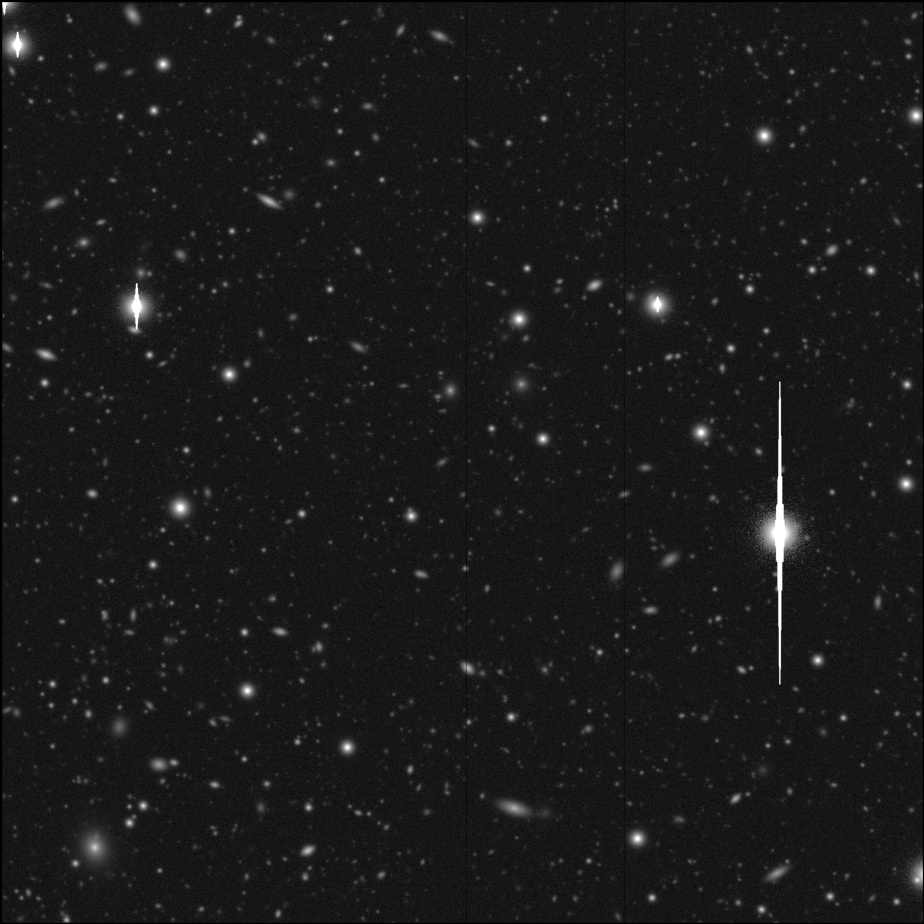}
    \end{subfigure}
    \caption{$2048{\times}2048$-pixel images simulated with the \texttt{descwl-shear-sims} package. Here, we display only the $r$ band of each image. The top-left panel displays an image from setting 2, which has 240 galaxies per square arcminute, a spatially varying PSF, no stars, and no image artifacts (\cref{subsec:psf}). The top-right panel displays an image from setting 3, which adds stars (\cref{subsec:stars}). The bottom-left panel displays an image from setting 4, which incorporates cosmic rays and bad CCD columns (\cref{subsec:artifacts}). The bottom-right panel displays an image from setting 5, which has a reduced raw density of 80 galaxies per square arcminute.}
    \label{fig:images}
\end{figure*}

We simulate cosmic rays and bad CCD columns using the procedure described in \cite{sheldon2023metadetection}. For each image in settings 4 and 5, cosmic ray positions are randomly sampled from a Poisson distribution. At each position, a cosmic ray is drawn with a uniformly random orientation and a length sampled uniformly between 10 and 30 pixels. In addition, a random number of bad columns is sampled from a Poisson distribution, and each is assigned a randomly selected position along the horizontal axis. Vertical gaps are randomly injected into some bad columns, with gap lengths uniformly distributed between 10\% and 30\% of the column length. The bottom-left panel of \cref{fig:images} illustrates a representative image from setting 4 that contains two bad columns. In \cref{subsec:train}, we describe how we handle the bad pixels introduced by these artifacts.

\section{Neural posterior estimation} \label{sec:npe}

\subsection{Notation and objective} \label{subsec:notationobjective}

Neural posterior estimation (NPE) is a simulation-based inference procedure for approximating $p(z \padvert x)$, the posterior distribution over latent random variables $z$ given observed random variables $x$. It involves training a neural network to infer the parameters $\phi$ of a variational distribution $q_\phi(z \padvert x)$ that approximates $p(z \padvert x)$.

In this work, we use NPE to infer shear from images simulated according to the implicit generative model of the \texttt{descwl-shear-sims} package. The latent vector of interest is $\gamma = [\gamma_1 \enspace \gamma_2]^\top$, which denotes the shear underlying a single multiband image. Here, $\gamma_1$ and $\gamma_2$ are treated as scalars and assumed to be constant within each image.

Let $z_{\text{nuisance}}$ denote all other latent variables encoded in the image, including the properties of the imaged galaxies, stars, pixel-level noise, detector artifacts, and PSF. Let $x$ denote the $3{\times}2048{\times}2048$ multiband image generated from $\gamma$ and $z_{\text{nuisance}}$ according to the forward model $p(x \padvert \gamma, z_{\text{nuisance}})$, which is defined implicitly based on the simulation settings described in \cref{sec:simulatedimages}.

Given an image $x$, we aim to infer $p(\gamma \padvert x)$, the posterior distribution over shear. Directly computing $p(\gamma \padvert x)$ is impractical because it requires integrating over the nuisance variables $z_{\text{nuisance}}$:
\begin{align}
    p(\gamma \padvert x) &= \int p(\gamma, z_{\text{nuisance}} \padvert x) \, dz_{\text{nuisance}} \\
    &= \int \frac{p(\gamma, z_{\text{nuisance}}) \, p(x \padvert \gamma, z_{\text{nuisance}})}{p(x)} \, dz_{\text{nuisance}}.
\end{align}

With NPE, we sidestep this intractability by approximating $p(\gamma \padvert x)$ with a variational distribution $q_\phi(\gamma \padvert x)$. In this work, we use a mean-field variational family \citep{zhang2019advances}, which approximates the joint posterior over the two shear components as a product of independent variational factors — i.e.,
\begin{equation} \label{eq:variationaldist}
q_\phi(\gamma \padvert x) = q_\phi(\gamma_1 \padvert x) q_\phi(\gamma_2 \padvert x),
\end{equation}
where each factor is an independent Gaussian. Thus, for each image, the vector of variational parameters $\phi$ comprises a mean and a variance for each of the two shear components.

NPE fits the variational distribution $q$ in an amortized fashion: we train a neural network $f_\eta$ to map each image $x$ to variational parameters $\phi$, such that $\phi = f_\eta(x)$. Notably, $x$ is the only input to $f_\eta$ in our implementation. While metadata such as the PSF could in principle be supplied as additional input, we found this to be unnecessary; the PSF and other nuisance variables enter the model solely through their effect on the training images.

The weights $\eta$ of the neural network are selected to minimize
\begin{align} \label{eq:npeobjective}
    \mathcal{L}(\eta) &= E_{p(x)} \left[D_{\text{KL}}\left(p(\gamma \padvert x) \thinspace\Vert\thinspace q_{f_\eta(x)} (\gamma \padvert x) \right)\right] \nonumber \\
    &= -E_{p(\gamma,x)}[\log q_{f_\eta(x)}(\gamma \padvert x)] + C,
\end{align}
the expected Kullback–Leibler divergence of the true posterior distribution from the variational distribution. In the second line, $C$ denotes a constant with respect to $\eta$. The expectation is evaluated over pairs $(\gamma, x)$ obtained by sampling from $p(\gamma, z_{\text{nuisance}}, x)$ and discarding the nuisance latent variables $z_{\text{nuisance}}$.

Provided their distributional form is sufficiently expressive, the variational factors $q_\phi(\gamma_1 \padvert x)$ and $q_\phi(\gamma_2 \padvert x)$ that minimize \cref{eq:npeobjective} coincide exactly with the true marginals $p(\gamma_1 \padvert x)$ and $p(\gamma_2 \padvert x)$, respectively \citep{ambrogioni2019forward}. In this work, we primarily assess the accuracy and calibration of the marginal posterior approximations for $\gamma_1$ and $\gamma_2$. While one could define a more flexible variational family, the mean-field Gaussian family is sufficient for our purposes.

\subsection{Neural network architecture} \label{subsec:architecture}

Our neural network must aggregate information from all $3{\times}2048{\times}2048$ pixels into a distributional summary of the constant shear vector $\gamma$. We define $f_{\eta}$ as a deep residual neural network (ResNet) \citep{he2016deep} that progressively downsamples the spatial dimensions of the input via several residual blocks. The residual blocks compress the spatial dimensions from $2048{\times}2048$ to $8{\times}8$ while simultaneously expanding and contracting the channel depth — the number of feature maps stored at each spatial location. The input image has three channels (one for each of the $r$, $i$, and $z$ bands), while the deepest residual blocks operate on tensors with up to 512 channels of learned features. The residual blocks are preceded by a convolutional preprocessing layer and followed by a fully connected layer that maps the flattened $8{\times}8$ feature maps to the variational parameters $\phi$. We describe the network architecture in greater detail in \cref{appdx:networkarchitecture}.

\section{Training and evaluation} \label{sec:traineval}

\subsection{Training the neural networks} \label{subsec:train}

We generate ten datasets of 5,000 images for each of the five simulation settings and randomly partition each one into training (3,000), validation (1,000), and test (1,000) sets. These replicate datasets allow us to characterize NPE's average performance and its variability under the stochasticity of image simulation, data splitting, and network training.

For each dataset, we train a separate network $f_\eta$ to minimize \cref{eq:npeobjective}. This results in ten independently trained networks per simulation setting. Training each $f_\eta$ takes approximately 60 hours on a single NVIDIA RTX 6000 GPU. We use a batch size of one, as we found that at most one $3{\times}2048{\times}2048$-pixel image fits in GPU memory along with the network weights.

Saturated stars with bleed trails introduce extreme pixel values in settings 3 through 5 (\cref{subsec:stars}), while cosmic rays and bad CCD columns introduce missing pixel values in settings 4 and 5 (\cref{subsec:artifacts}). To limit the dynamic range introduced by saturated stars, we clamp pixel values at the 99th percentile of each multiband image before feeding them to the network. We impute missing pixel values with the median pixel value of the image.

\subsection{Shear estimation benchmark: AnaCal} \label{subsec:anacal}

We compare NPE to the AnaCal shear estimation algorithm, which we view as a natural benchmark since it has previously been tested on \texttt{descwl-shear-sims} images \citep{li2023analytical, li2024differentiable, li2025analytical}. We benchmark against AnaCal but not Metadetection primarily for computational convenience. See \cite{li2025analytical} for a detailed derivation and algorithmic description of AnaCal.

We run AnaCal using the \texttt{anacal} package.\footnote{\url{https://github.com/mr-superonion/AnaCal}} We encountered difficulties replicating the masking procedure used by AnaCal to account for stars and image artifacts, so we apply the algorithm only to settings 1 and 2. For each of these settings, we apply AnaCal to the same ten 1,000-image test sets as NPE, using 28 parallel workers on 36 CPU cores.

The PSF is a required input to AnaCal. For images with a fixed PSF (setting 1), we render the PSF as a $64{\times}64$-pixel stamp and pass this stamp to AnaCal. For images with a spatially varying PSF (setting 2), we pass the \texttt{GalSim} PSF object directly to AnaCal.\footnote{\url{https://github.com/GalSim-developers/GalSim}}

To run AnaCal on the multiband images, we first combine the $r$, $i$, and $z$ exposures into a single-band image using inverse-variance weighting \citep{li2025analytical}. Here, the variances are derived from the projected year-10 LSST noise in each band.

For each image, AnaCal produces a point estimate for each shear component. Following \cite{li2025analytical}, we estimate the shear response for each test set by pooling measurements across all images in the test set rather than estimating it separately for each image.

\subsection{Evaluation metrics and plots} \label{subsec:eval}

We treat the means of the variational distributions inferred by NPE as point estimates for shear, which facilitates comparison with AnaCal. We evaluate the NPE and AnaCal point estimates using three metrics, which we compute on each of the ten test sets of $N{=}1{,}000$ images for each simulation setting.

First, we estimate the multiplicative and additive biases of the two algorithms' estimates. We use the conventional parametrization
\begin{equation} \label{eq:multiplicativebias}
    \widehat{\gamma}_j^n = (1 + m_j) \gamma_j^n + c_j
\end{equation}
for $n \in \{1,\ldots,N\}$ and $j \in \{1,2\}$, where $m_j$ is the multiplicative bias and $c_j$ is the additive bias \citep{heymans2006shear, huterer2006systematic}. For each test set, we estimate $(1+m_j)$ and $c_j$ via ordinary least squares.

We also compute the root mean squared error (RMSE) and Pearson correlation coefficient ($r$). For $j \in \{1,2\}$, the former is computed for each test set as
\begin{equation} \label{eq:rmse}
    \text{RMSE}\left(\gamma_j, \widehat{\gamma}_j\right) = \sqrt{\frac{1}{N} \sum_{n=1}^N \left(\gamma_j^n - \widehat{\gamma}_j^n\right)^2},
\end{equation}
and the latter is
\begin{equation} \label{eq:pearson}
    r\left(\gamma_j, \widehat{\gamma}_j\right) = \frac{\text{Cov}\left(\gamma_j, \widehat{\gamma}_j\right)}{\sqrt{\text{Var}\left(\gamma_j\right) \text{Var}\left(\widehat{\gamma}_j\right)}},
\end{equation}
where $\text{Cov}(\cdot,\cdot)$ denotes the sample covariance of the input vectors and $\text{Var}(\cdot)$ denotes the sample variance. In \cref{table:metrics}, we report the mean and standard deviation of the multiplicative biases, RMSEs, and Pearson correlation coefficients across the ten test sets for each setting. We also report the average per-image inference time of NPE (post-training) and AnaCal for each setting in \cref{appdx:timingcomparison}.

Since distilling the variational distributions inferred by NPE to point estimates is reductive, we also assess the calibration of credible intervals derived from these variational distributions. For one randomly selected test set per simulation setting, we plot 90\% highest-density credible intervals (NPE) and point estimates (AnaCal) against the true shears for the 1,000 test images.

We additionally construct credible intervals for nominal coverage levels between 0.05 and 0.95. For each nominal coverage level, we compute the empirical coverage as the proportion of the test images (pooled across the ten test sets for each setting) for which the credible interval covers the ground truth for each shear component. We plot these empirical coverages against the corresponding nominal coverages. The points on this plot will lie close to the diagonal if the credible intervals are well-calibrated, above the diagonal if they are systematically too wide, and below the diagonal if they are systematically too narrow. These coverage plots, which are similar in spirit to simulation-based calibration \citep{cook2006validation, talts2018validating}, are commonly used to evaluate variational inference algorithms (e.g., \citealt{cannon2022investigating, duan2026neural}).

Finally, we visualize the variational densities inferred by NPE for several representative test images in each simulation setting (\cref{appdx:densityplots}). On each plot, we overlay the true shears; for settings 1 and 2, we also overlay the AnaCal point estimates.

\begin{table*}
    \centering
    \begin{tabular}{ll ccc ccc}
        \toprule
        & & \multicolumn{3}{c}{\textbf{AnaCal}} & \multicolumn{3}{c}{\textbf{NPE}} \\
        \cmidrule(r){3-5} \cmidrule(l){6-8}
        & & $m$ & RMSE & Pearson $r$ & $m$ & RMSE & Pearson $r$ \\
        \midrule
        \multirow{2}{*}{\textbf{Setting 1}} & $\gamma_1$ & 0.002 (0.014) & 0.0065 (0.0002) & 0.922 (0.005) & -0.012 (0.015) & 0.0014 (0.0002) & 0.996 (0.001) \\
                                   & $\gamma_2$ & -0.008 (0.015) & 0.0070 (0.0004) & 0.911 (0.005) & -0.007 (0.016) & 0.0014 (0.0002) & 0.996 (0.002) \\
        \midrule
        \multirow{2}{*}{\textbf{Setting 2}} & $\gamma_1$ & 0.008 (0.018) & 0.0073 (0.0003) & 0.904 (0.005) & -0.015 (0.016) & 0.0016 (0.0002) & 0.995 (0.001) \\
                                   & $\gamma_2$ & 0.003 (0.015) & 0.0078 (0.0002) & 0.890 (0.009) & -0.012 (0.011) & 0.0017 (0.0002) & 0.994 (0.002) \\
        \midrule
        \multirow{2}{*}{\textbf{Setting 3}} & $\gamma_1$ & --- & --- & --- & -0.027 (0.018) & 0.0018 (0.0002) & 0.993 (0.001) \\
                                   & $\gamma_2$ & --- & --- & --- & -0.031 (0.017) & 0.0019 (0.0003) & 0.993 (0.002) \\
        \midrule
        \multirow{2}{*}{\textbf{Setting 4}} & $\gamma_1$ & --- & --- & --- & -0.032 (0.012) & 0.0018 (0.0002) & 0.993 (0.001) \\
                                   & $\gamma_2$ & --- & --- & --- & -0.034 (0.013) & 0.0018 (0.0001) & 0.993 (0.001) \\
        \midrule
        \multirow{2}{*}{\textbf{Setting 5}} & $\gamma_1$ & --- & --- & --- & -0.026 (0.020) & 0.0018 (0.0002) & 0.994 (0.001) \\
                                   & $\gamma_2$ & --- & --- & --- & -0.031 (0.018) & 0.0019 (0.0003) & 0.994 (0.001) \\
        \bottomrule
    \end{tabular}%
    \caption{Multiplicative biases (\cref{eq:multiplicativebias}), root mean squared errors (\cref{eq:rmse}), and Pearson correlation coefficients (\cref{eq:pearson}) of AnaCal and NPE shear estimates. Each column reports the mean of the metric across the ten test sets, with the standard deviation in parentheses. For AnaCal, the metrics are computed using the algorithm's point estimate for each image. For NPE, they are computed using the means of the inferred variational distributions.}
    \label{table:metrics}
\end{table*}

\section{Results}\label{sec:experiments}

We evaluate NPE's performance across the five simulation settings, using AnaCal as a benchmark for the first two as described in \cref{subsec:anacal}.

\subsection{Setting 1} \label{subsec:setting1}

\begin{figure*}
    \centering
    \begin{subfigure}[c]{0.48\textwidth}
        \includegraphics[width=\textwidth]{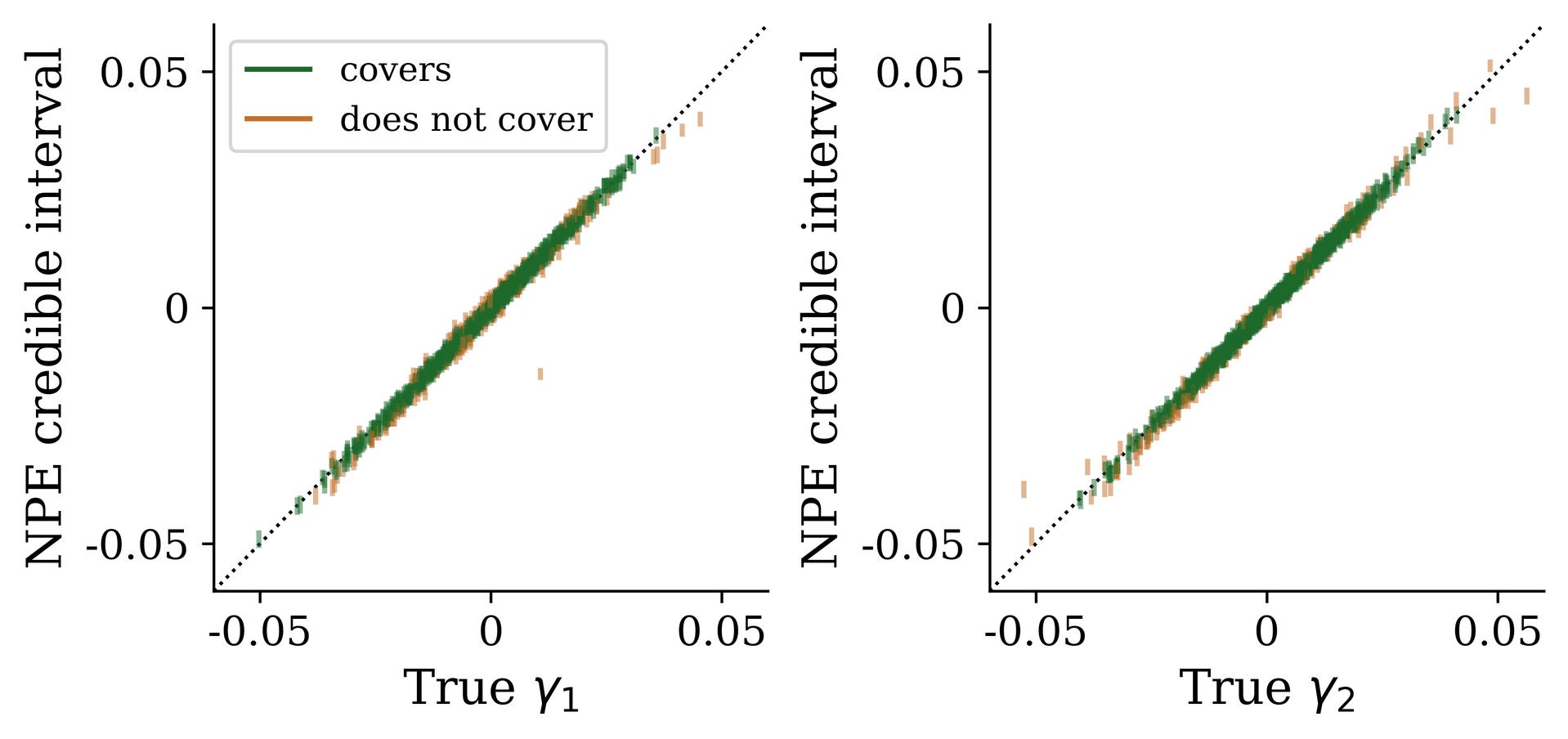}
    \end{subfigure}
    \hfill
    \begin{subfigure}[c]{0.48\textwidth}
        \includegraphics[width=\textwidth]{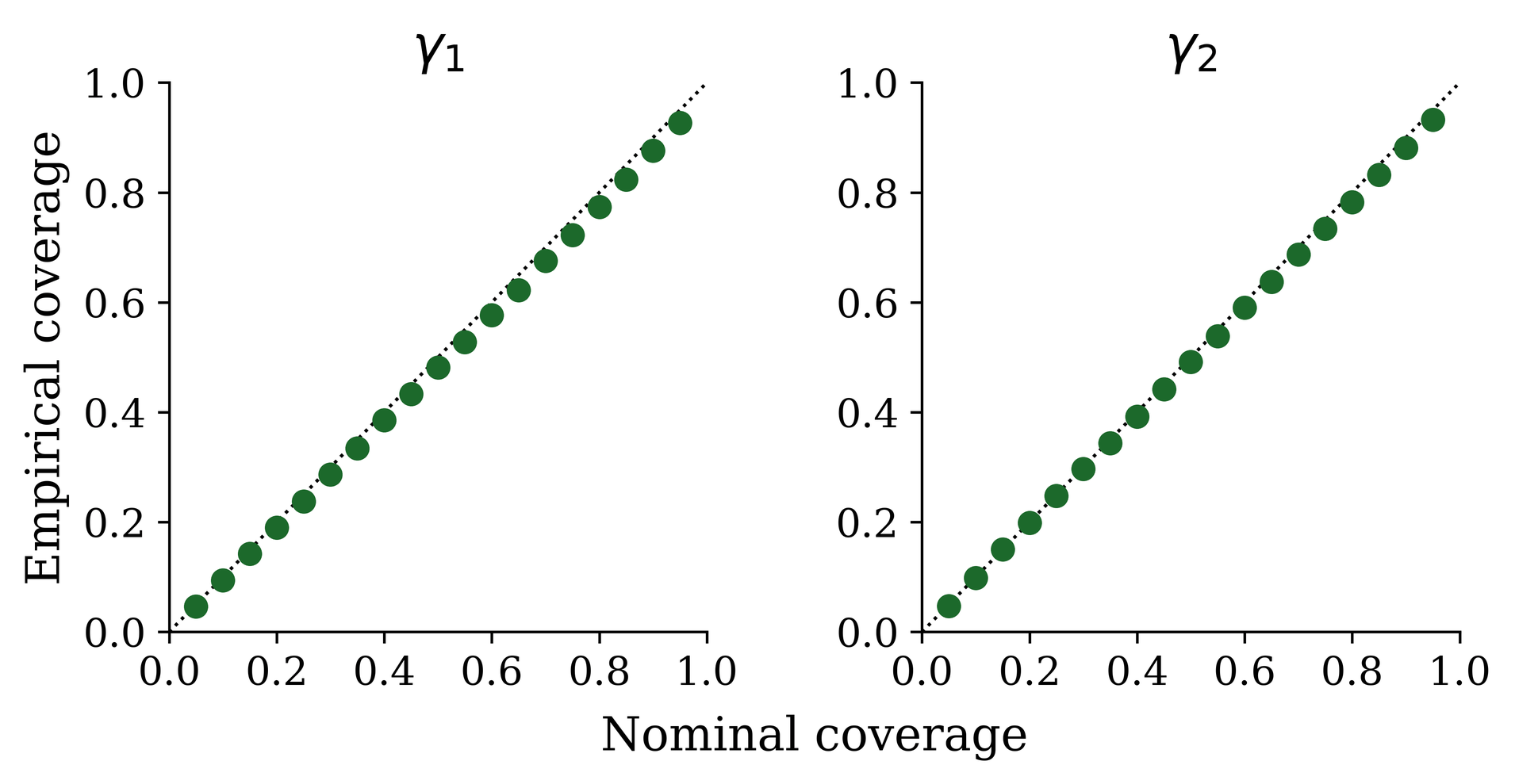}
    \end{subfigure}
    \caption{NPE calibration plots for setting 1. Left: True $\gamma_1$ and $\gamma_2$ versus 90\% credible intervals inferred by NPE for a randomly selected 1,000-image test set. Green intervals cover the true shear (dotted line), while orange intervals do not. Right: Nominal versus empirical coverage of credible intervals for $\gamma_1$ and $\gamma_2$, computed by pooling across the ten test sets.}
    \label{fig:credibleintervals1}
\end{figure*}

For the baseline images with a fixed PSF, no stars, and no detector artifacts, NPE produces well-calibrated posterior approximations for shear. The left panel of \cref{fig:credibleintervals1} plots 90\% credible intervals against the corresponding true shears; approximately 90\% of these intervals cover the ground truth for each shear component. The right panel of \cref{fig:credibleintervals1} demonstrates that the empirical coverage closely tracks the nominal coverage across all levels between 0.05 and 0.95.

\begin{figure*}
    \centering
    \begin{subfigure}[c]{0.48\textwidth}
        \includegraphics[width=\textwidth]{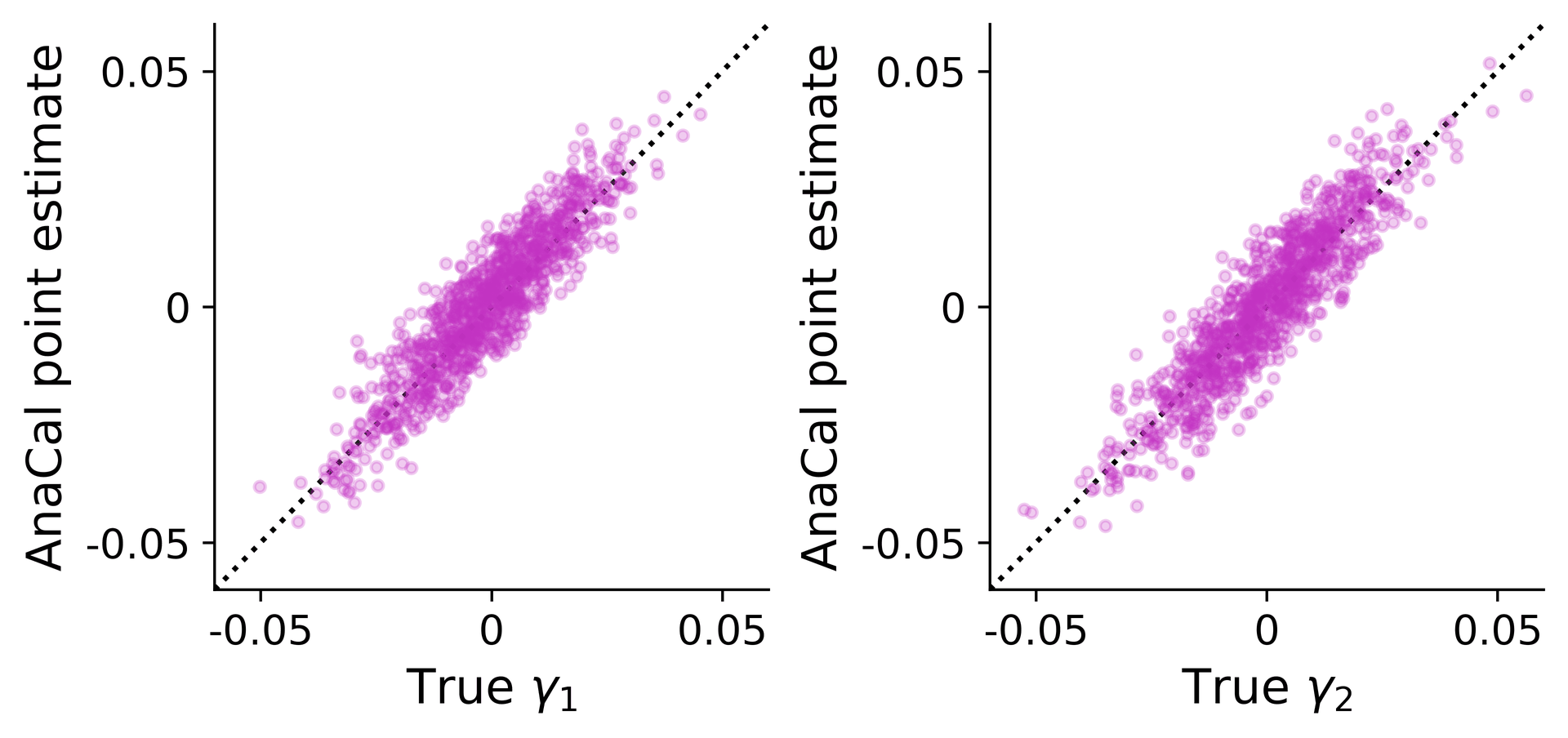}
    \end{subfigure}
    \hfill
    \begin{subfigure}[c]{0.48\textwidth}
        \includegraphics[width=\textwidth]{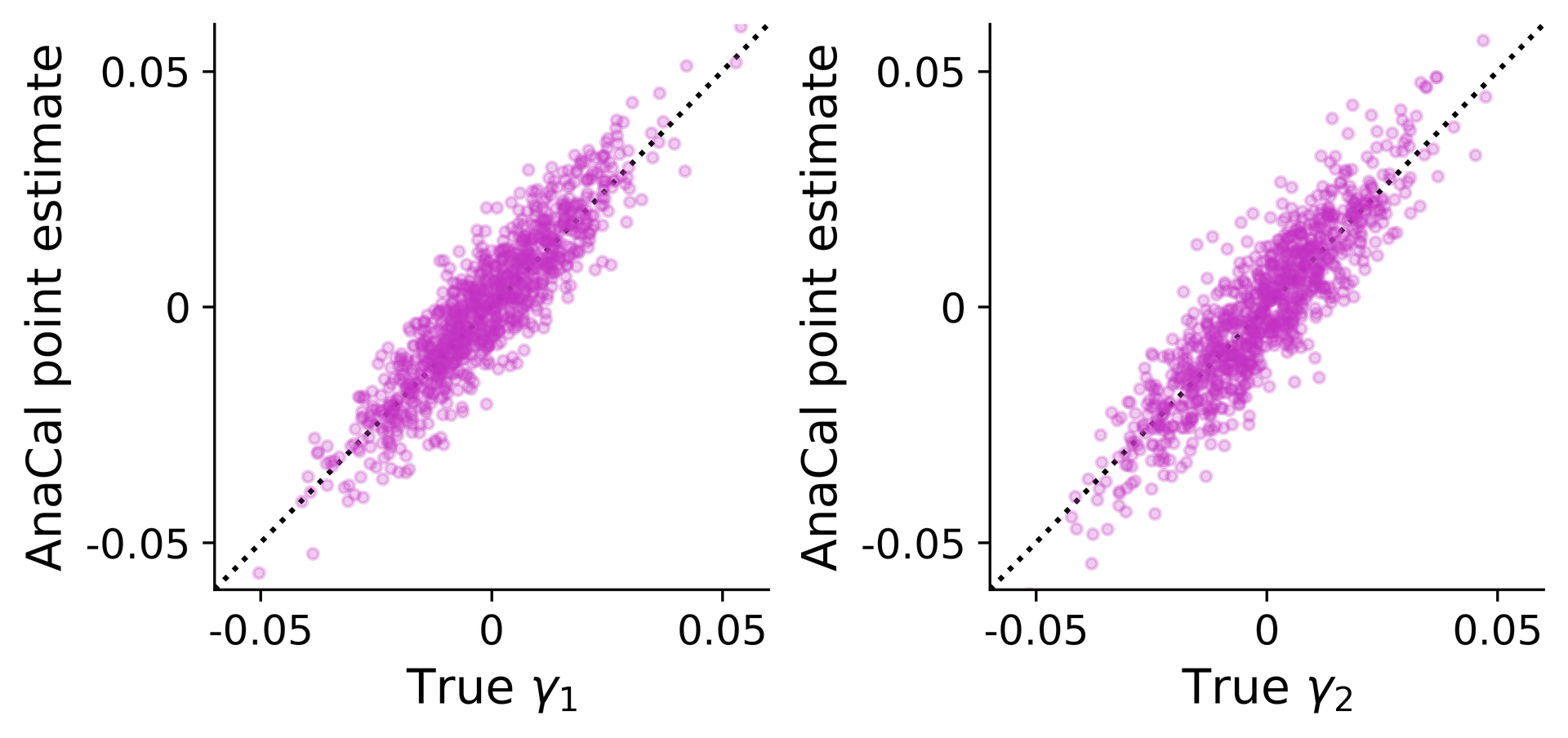}
    \end{subfigure}
    \caption{AnaCal scatterplots for settings 1 and 2. True $\gamma_1$ and $\gamma_2$ versus AnaCal point estimates for a randomly selected 1,000-image test set from setting 1 (left) and setting 2 (right).}
    \label{fig:anacal12}
\end{figure*}

As shown in \cref{table:metrics}, the NPE posterior means achieve lower RMSEs and higher Pearson correlation coefficients than the AnaCal point estimates, while AnaCal achieves a smaller multiplicative bias. This trade-off is evident in the left panels of \cref{fig:credibleintervals1} and \cref{fig:anacal12}: whereas the AnaCal estimates are centered along the 45-degree line but show greater vertical spread, the centers of the NPE credible intervals track the 45-degree line tightly but deviate slightly at either extreme. The two methods produce inferences at comparable per-image speeds in this setting (\cref{table:timing}).

\subsection{Setting 2} \label{subsec:setting2}

In our NPE implementation, the neural network does not use any information about the spatially varying PSF model introduced in setting 2. The key question in this section is whether NPE can account for PSF variation implicitly and retain its calibration and accuracy — particularly its advantage in RMSE and Pearson correlation over AnaCal, which takes the \texttt{GalSim} PSF object as an input.

\begin{figure*}
    \centering
    \begin{subfigure}[c]{0.48\textwidth}
        \includegraphics[width=\textwidth]{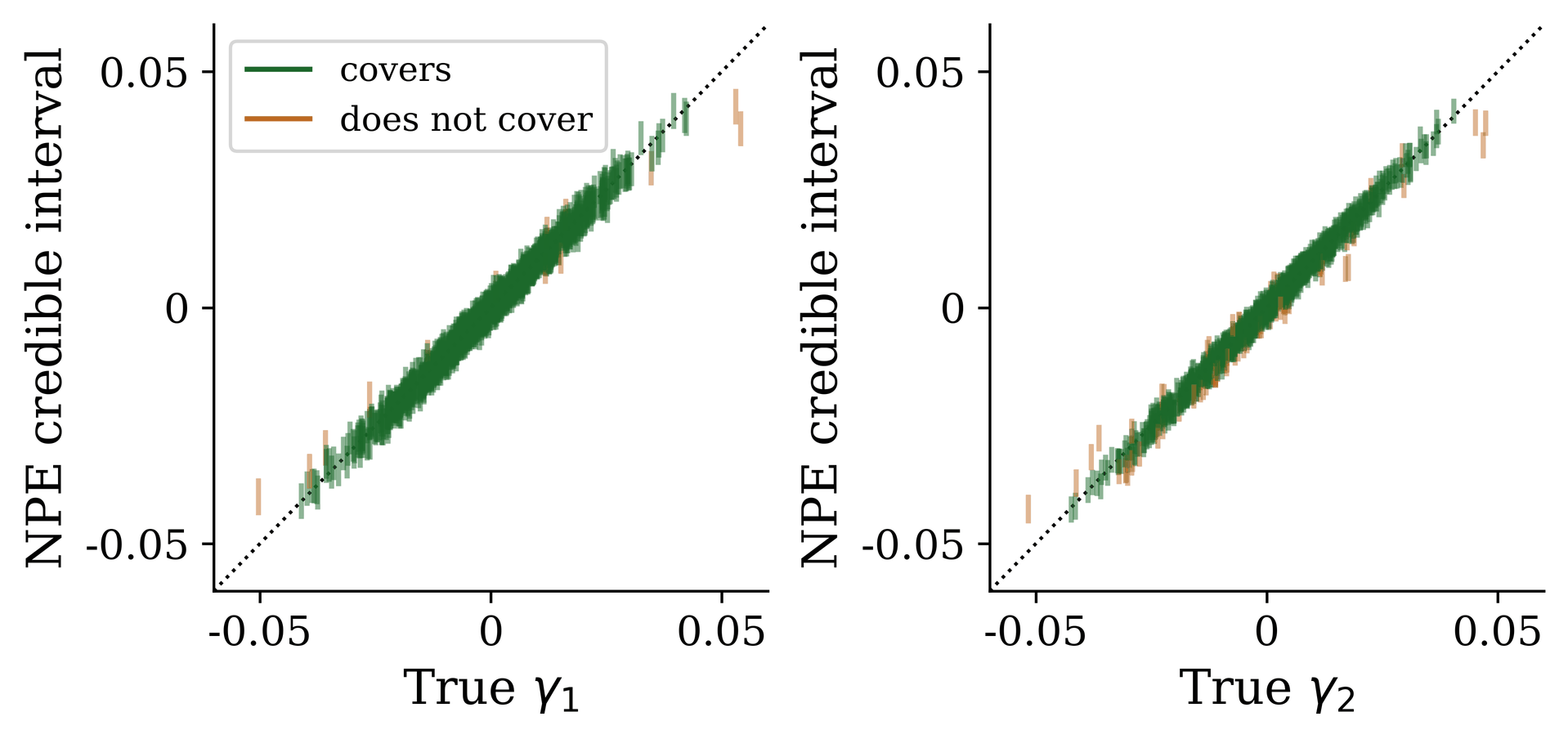}
    \end{subfigure}
    \hfill
    \begin{subfigure}[c]{0.48\textwidth}
        \includegraphics[width=\textwidth]{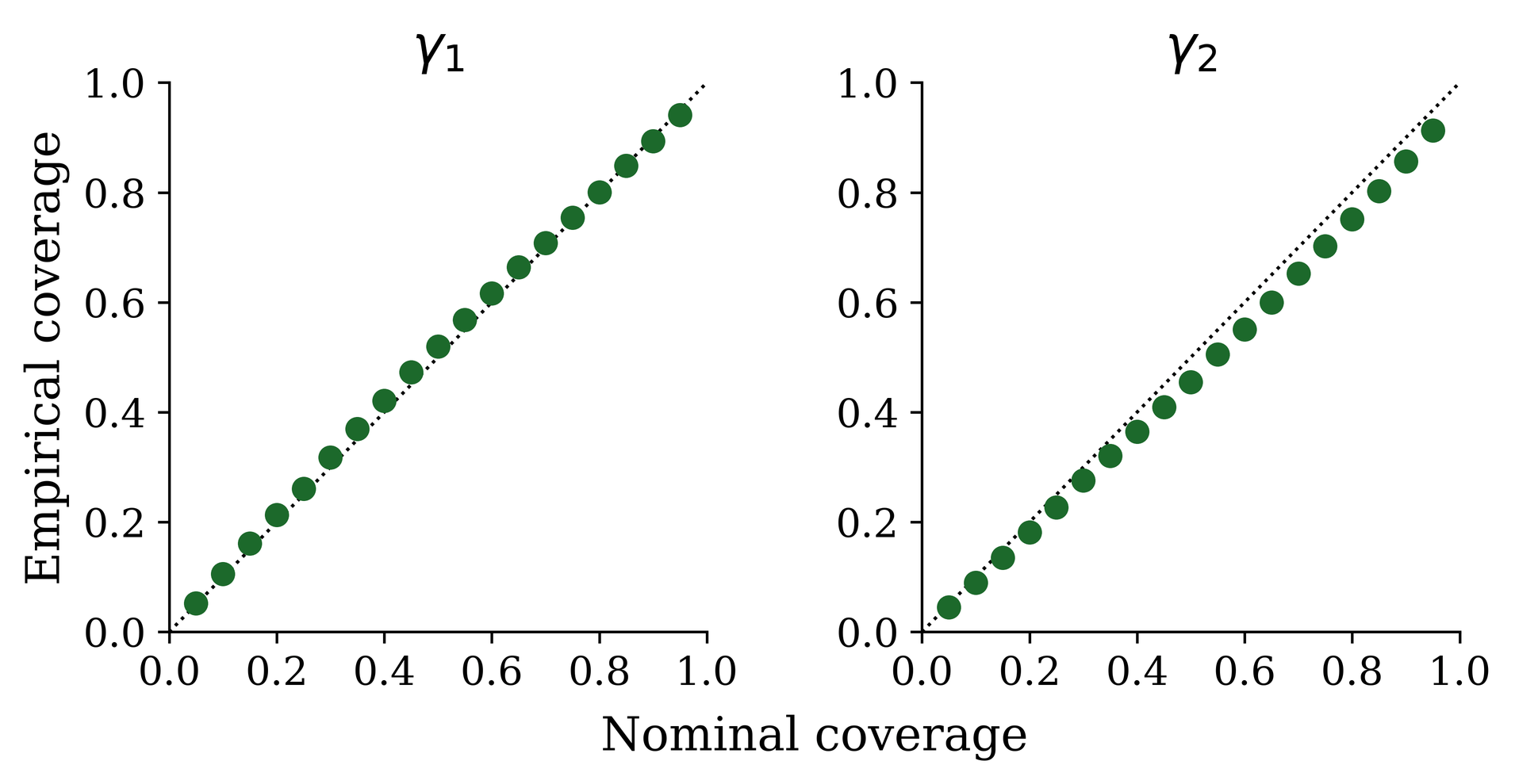}
    \end{subfigure}
    \caption{NPE calibration plots for setting 2. Left: True $\gamma_1$ and $\gamma_2$ versus 90\% credible intervals inferred by NPE for a randomly selected 1,000-image test set. Green intervals cover the true shear (dotted line), while orange intervals do not. Right: Nominal versus empirical coverage of credible intervals for $\gamma_1$ and $\gamma_2$, computed by pooling across the ten test sets.}
    \label{fig:credibleintervals2}
\end{figure*}

\Cref{fig:credibleintervals2} indicates that NPE produces well-calibrated posterior approximations for shear in setting 2. \Cref{table:metrics} suggests that inferring shear is slightly more challenging for NPE in this setting, as it achieves larger RMSEs and smaller Pearson correlation coefficients than in setting 1. However, the magnitude of these changes is small. These results provide evidence that NPE can handle spatially varying PSFs implicitly, without using them as inputs.

As in setting 1, NPE achieves lower RMSEs and higher Pearson correlation coefficients than AnaCal, while AnaCal achieves a smaller multiplicative bias for both shear components. We caution against comparing the multiplicative biases reported in \cref{table:metrics} to the LSST science requirements threshold \citep{mandelbaum2018lsst}, as our test sets cover fewer square degrees than those used in related works where the multiplicative bias is the primary evaluation metric (e.g., \citealt{li2025analytical}).

AnaCal is several times slower than NPE in this setting (\cref{table:timing}). This slowdown is a byproduct of our AnaCal implementation, which constructs the spatially varying \texttt{GalSim} PSF object for each image and evaluates it at each galaxy position. This overhead could potentially be avoided; for example, \citet{li2025analytical} approximate the PSF as constant within coarse spatial cells.

\subsection{Setting 3} \label{subsec:setting3}

The setting 3 images contain stars, including saturated stars with bleed trails. Our NPE implementation does not detect and mask these stars as a preprocessing step; we address the extreme pixel values using only the basic clipping procedure described in \cref{subsec:train}.

\begin{figure*}
    \centering
    \begin{subfigure}[c]{0.48\textwidth}
        \includegraphics[width=\textwidth]{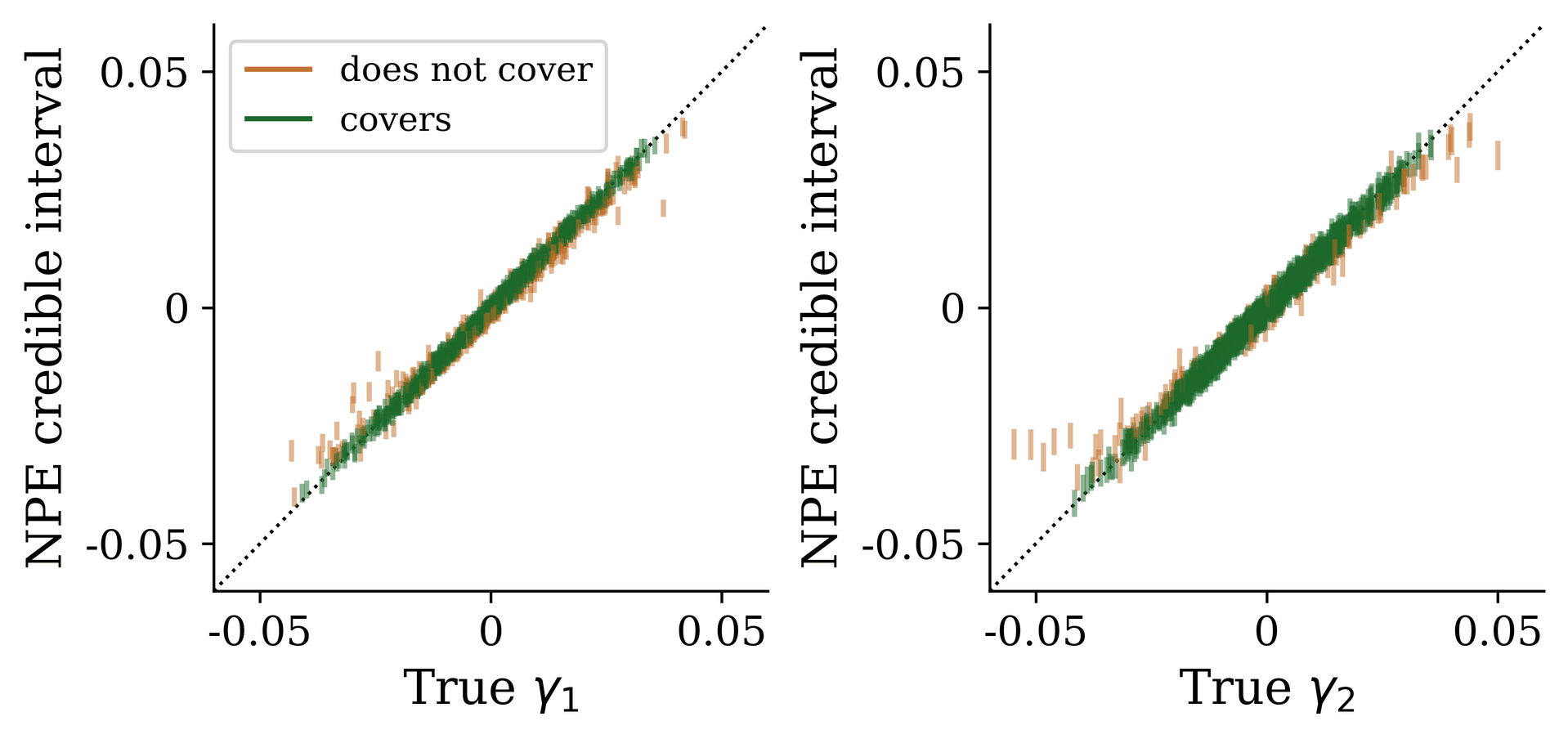}
    \end{subfigure}
    \hfill
    \begin{subfigure}[c]{0.48\textwidth}
        \includegraphics[width=\textwidth]{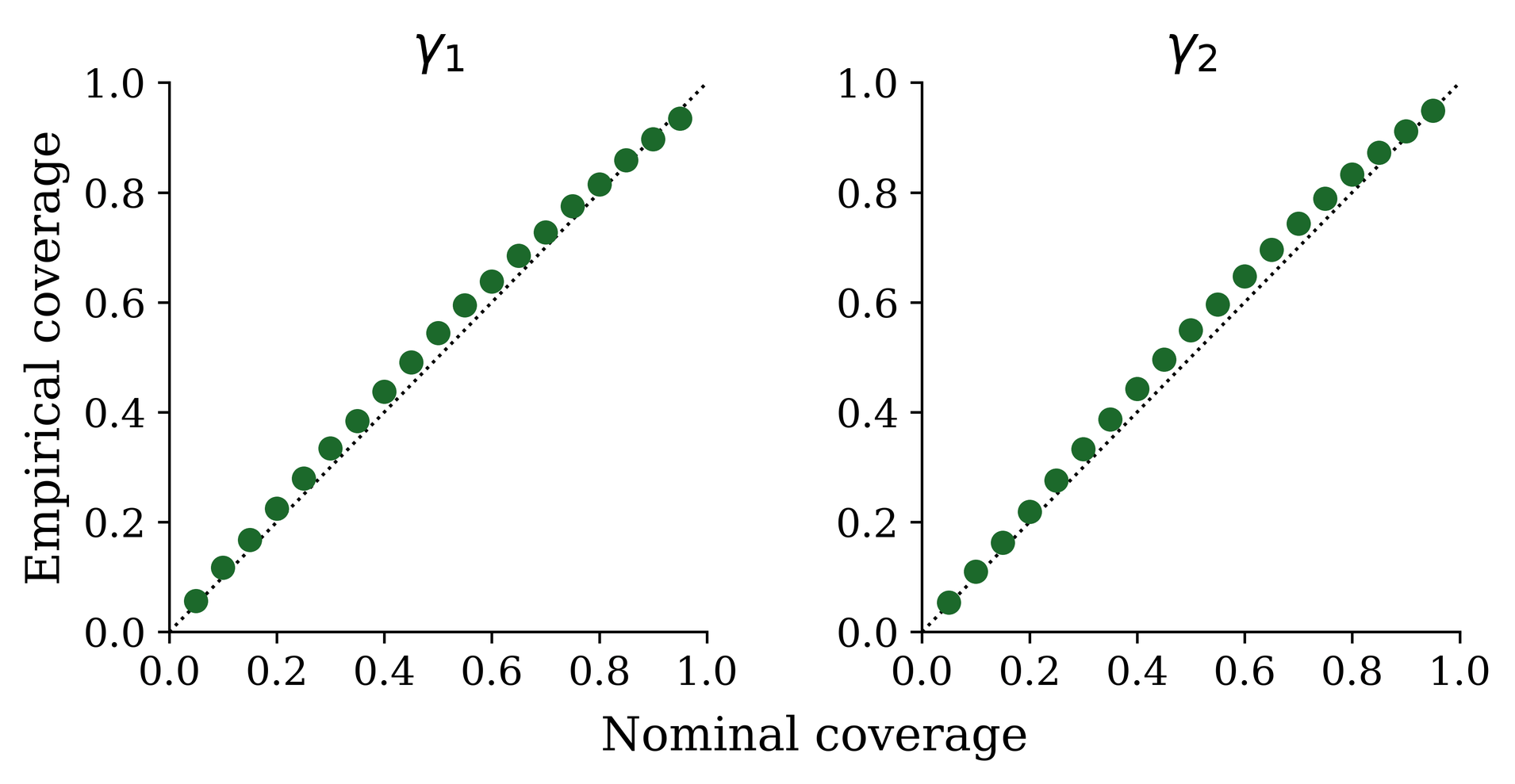}
    \end{subfigure}
    \caption{NPE calibration plots for setting 3. Left: True $\gamma_1$ and $\gamma_2$ versus 90\% credible intervals inferred by NPE for a randomly selected 1,000-image test set. Green intervals cover the true shear (dotted line), while orange intervals do not. Right: Nominal versus empirical coverage of credible intervals for $\gamma_1$ and $\gamma_2$, computed by pooling across the ten test sets.}
    \label{fig:credibleintervals3}
\end{figure*}

In this setting, the NPE posterior means have multiplicative biases that are larger in magnitude than in settings 1 and 2, as well as slightly larger RMSEs and slightly smaller Pearson correlation coefficients. The right panel of \cref{fig:credibleintervals3} indicates that the NPE credible intervals remain well-calibrated for both shear components, with empirical coverage tracking nominal coverage closely across all levels. These results suggest that NPE can implicitly handle stellar contamination without explicit masking.

\subsection{Setting 4} \label{subsec:setting4}

\begin{figure*}
    \centering
    \begin{subfigure}[c]{0.48\textwidth}
        \includegraphics[width=\textwidth]{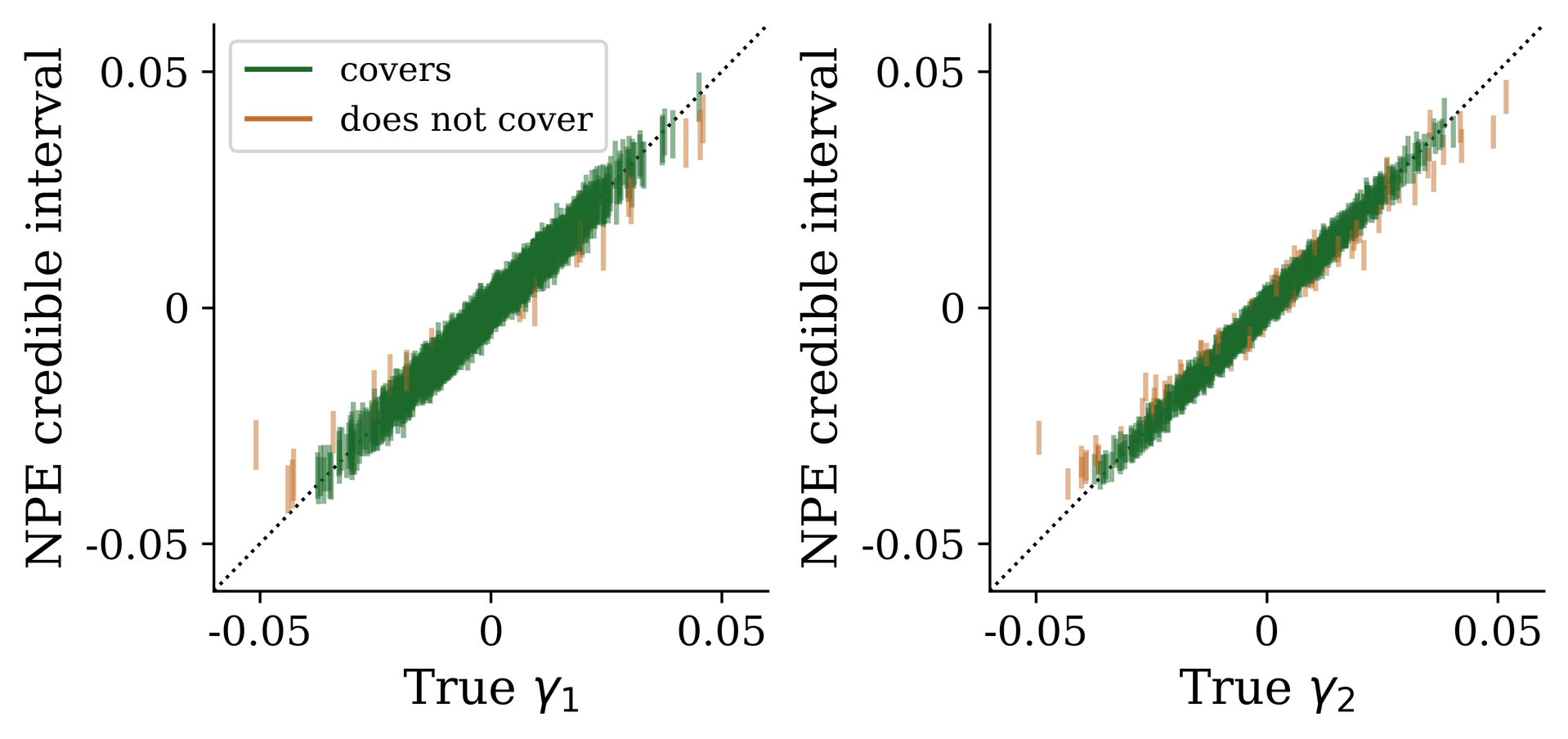}
    \end{subfigure}
    \hfill
    \begin{subfigure}[c]{0.48\textwidth}
        \includegraphics[width=\textwidth]{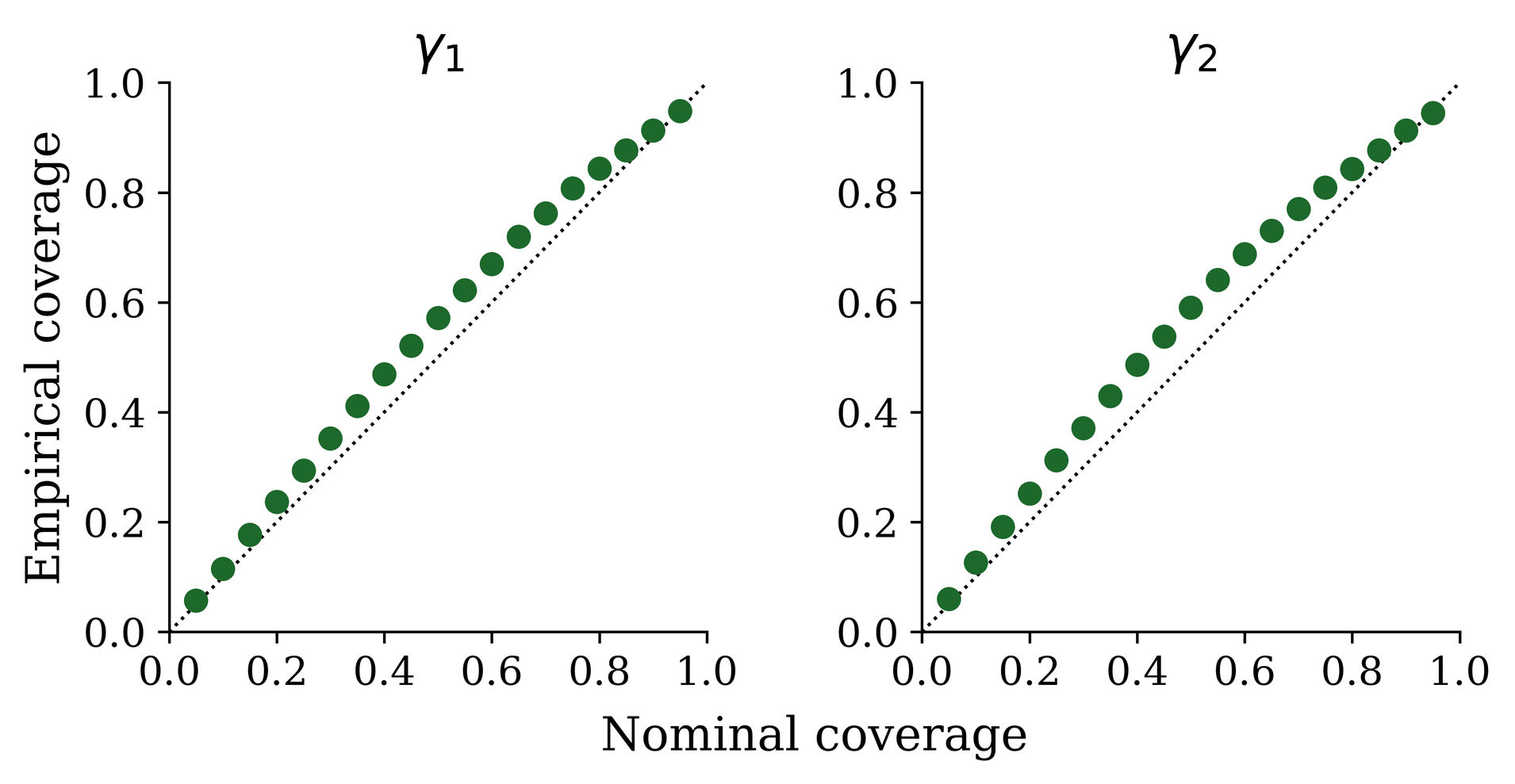}
    \end{subfigure}
    \caption{NPE calibration plots for setting 4. Left: True $\gamma_1$ and $\gamma_2$ versus 90\% credible intervals inferred by NPE for a randomly selected 1,000-image test set. Green intervals cover the true shear (dotted line), while orange intervals do not. Right: Nominal versus empirical coverage of credible intervals for $\gamma_1$ and $\gamma_2$, computed by pooling across the ten test sets.}
    \label{fig:credibleintervals4}
\end{figure*}

Based on all three metrics reported in \cref{table:metrics}, the accuracy of the NPE posterior means is not noticeably impacted by the cosmic rays and bad CCD columns introduced in setting 4. The right panel of \cref{fig:credibleintervals4} indicates that the NPE credible intervals tend to be slightly too wide for both shear components, as the empirical coverages are larger than the nominal coverages by a few percentage points for most levels. The 90\% credible intervals in the left panel are slightly wider than in previous settings.

In general, however, NPE's posterior approximations remain accurate and well-calibrated in the presence of cosmic rays and bad columns. NPE requires the missing pixel values arising from these detector artifacts to be imputed before training, but our results imply that the relatively simple median imputation procedure described in \cref{subsec:train} is sufficient.

\begin{figure*}
    \centering
    \begin{subfigure}[c]{0.48\textwidth}
        \includegraphics[width=\textwidth]{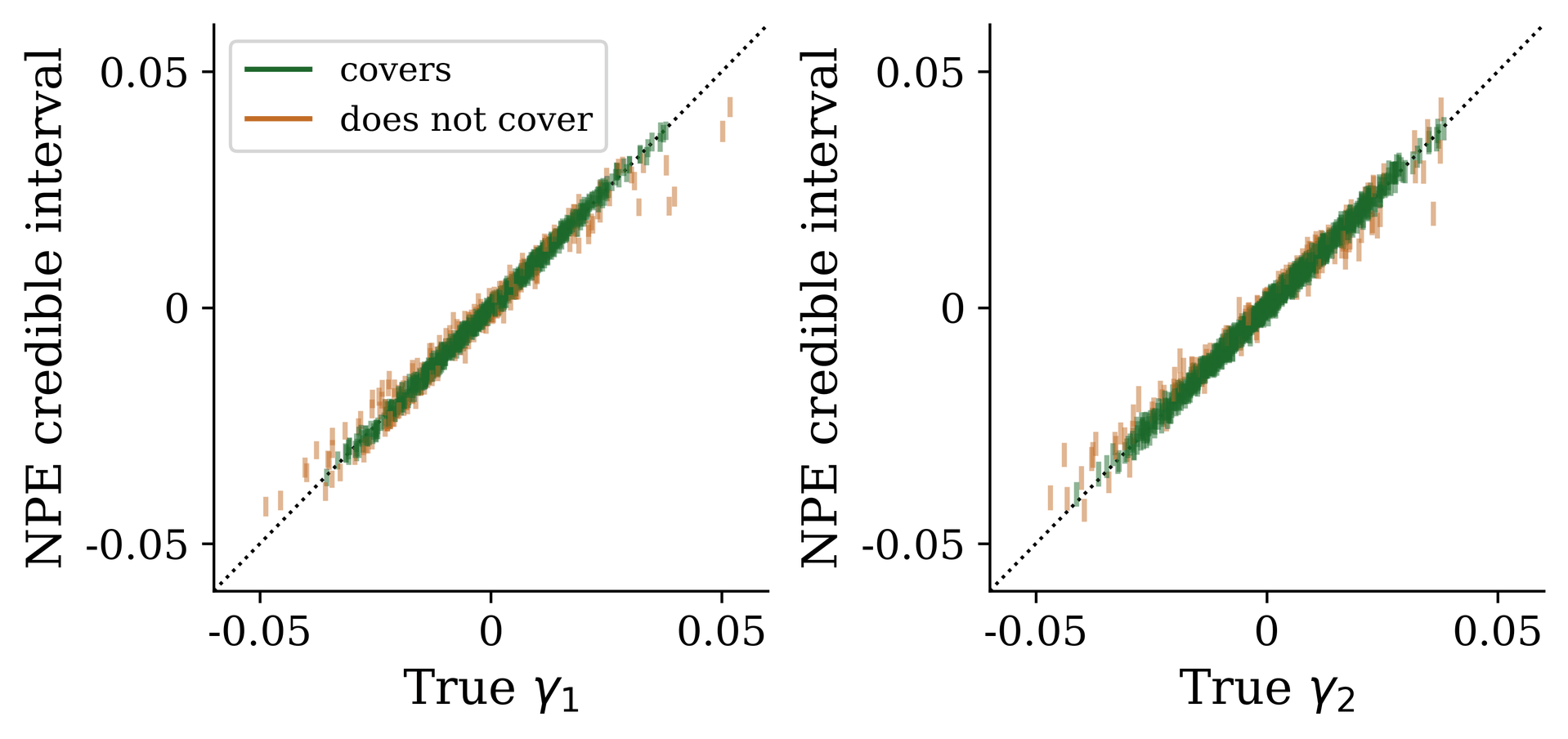}
    \end{subfigure}
    \hfill
    \begin{subfigure}[c]{0.48\textwidth}
        \includegraphics[width=\textwidth]{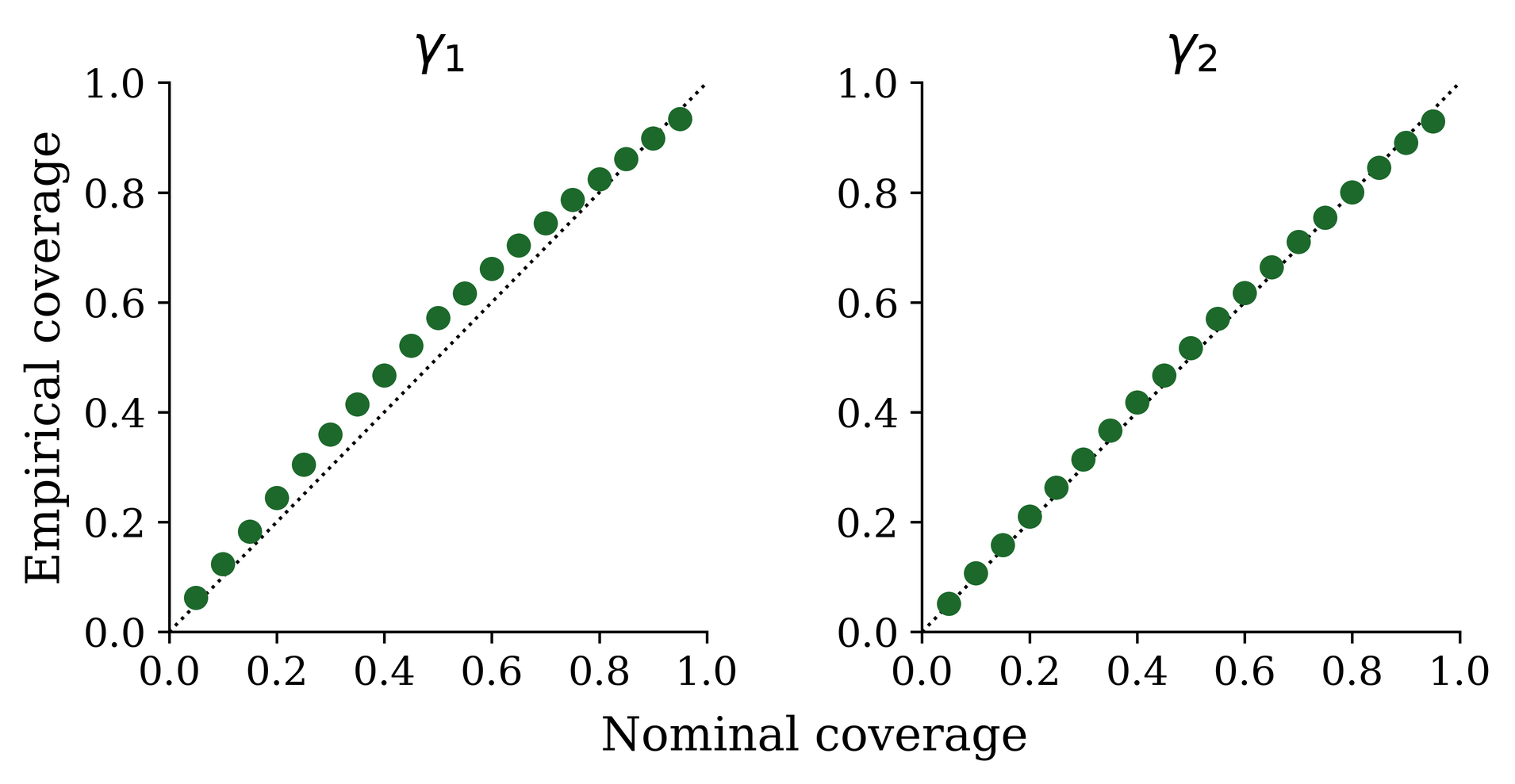}
    \end{subfigure}
    \caption{NPE calibration plots for setting 5. Left: True $\gamma_1$ and $\gamma_2$ versus 90\% credible intervals inferred by NPE for a randomly selected 1,000-image test set. Green intervals cover the true shear (dotted line), while orange intervals do not. Right: Nominal versus empirical coverage of credible intervals for $\gamma_1$ and $\gamma_2$, computed by pooling across the ten test sets.}
    \label{fig:credibleintervals5}
\end{figure*}

\subsection{Setting 5} \label{subsec:setting5}

The final simulation setting reduces the galaxy density from 240 to 80 galaxies per square arcminute. This presents a trade-off, as the sparser images provide less information to the network about the underlying shear field, but the decreased blending in these images makes it easier for the network to measure individual galaxy shapes.

\Cref{table:metrics} shows that the NPE posterior means achieve multiplicative biases, RMSEs, and Pearson correlation coefficients comparable to those in settings 3 and 4, indicating that the reduced galaxy density does not substantially degrade the accuracy of these point estimates. \Cref{fig:credibleintervals5} demonstrates that the NPE credible intervals remain well-calibrated for both shear components.

Across all five simulation settings, NPE's multiplicative biases are negative, which reflects the networks' tendency to underestimate shear in images where it is large in magnitude. This may be a consequence of our Gaussian prior distribution, which produces relatively fewer samples at high-amplitude shear. In future work, we plan to investigate whether switching to a uniform prior distribution would improve NPE's accuracy on images with high-amplitude shear.

\section{Discussion}\label{sec:discussion}

In experiments on several sets of simulated constant-shear images with increasingly complex observational effects, our neural posterior estimation (NPE) procedure produces accurate and well-calibrated variational distributions for shear. Our results indicate that a neural network trained to minimize the NPE objective can learn to marginalize over blended galaxies, spatially varying point spread functions, stars, and detector artifacts when (i) these effects are present in the training images and (ii) the training and evaluation images are generated by a common simulator.

Our algorithm differs fundamentally from conventional shear estimators, which account for these potential sources of bias primarily by calibrating estimates of galaxy ellipticities. NPE shifts the burden of systematic bias control to the forward model, a strategy that will become increasingly appealing as simulation fidelity continues to improve \citep{stone2023astrophot, openuniverse2025}.

\subsection{Sensitivity to model misspecification} \label{subsec:limitations}

In our experiments, the training and evaluation images are generated by the same probabilistic model, which is specified implicitly by the \texttt{descwl-shear-sims} package. This setup enables us to isolate each additional observational effect without conflating its impact with distribution shift. However, our simulations rely on several simplifications: the pixel-level noise in the coadd coordinate frame is independent and Gaussian, all galaxies are rendered at the same redshift, the shear field is constant within each image, and convergence is set to zero. The latter three simplifications stem from the limitations of \texttt{descwl-shear-sims}, which does not currently support spatially varying shear and convergence fields or galaxy catalogs spanning multiple redshifts. Once higher-fidelity simulators become available, we plan to assess the performance of NPE under correlated pixel-level noise and tomographic, spatially varying shear and convergence, perhaps using an incremental experimental setup similar to the one in this work.

In practice, NPE would be trained on simulated images and applied to real astronomical surveys whose data-generating process is only approximately captured by the simulator. Discrepancies between the generative models of the simulator and the survey can lead to biased or miscalibrated posterior approximations \citep{hermans2022crisis, frazier2024statistical}. Addressing model misspecification in simulation-based inference is an active area of research, with some works proposing methods for diagnosing misspecification \citep{schmitt2024detecting, chen2025colt} and others proposing methods for mitigating it (see \citealt{kelly2025simulation} for a recent review). We plan to investigate the impact of model misspecification on our shear estimation procedure in future work.

\subsection{Connection to other NPE applications}

The present study builds on our collaboration's previous applications of NPE to astronomical image processing. Our findings align with those of \citet{patel2025neural}, who use NPE for source detection, classification, and measurement in images with spatially varying backgrounds and PSFs. They demonstrate that withholding explicit PSF information does not degrade NPE's performance on these tasks, which suggests that the network can implicitly learn the distribution of PSFs from stars and galaxies in the training images. Our results for simulation settings 2-5 likewise show that NPE accurately infers shear in images with spatially varying PSFs, even when the network receives no explicit information about the PSF model.

This work also complements the approach proposed by \citet{white2026tomographic}, which uses NPE to approximate the posterior distribution over tomographic shear and convergence maps — i.e., spatially varying maps of shear and convergence for several redshift bins, represented as three-dimensional arrays. That paper applies NPE to the LSST-DESC DC2 Simulated Sky Survey. The DC2 images were generated by a computationally intensive simulator that was run once, thus precluding direct control over individual observational effects. In contrast, the modular \texttt{descwl-shear-sims} simulator used here enables incremental introduction of specific effects. Combining this controlled sensitivity analysis with the more complex inference task in \citet{white2026tomographic} would provide additional insight into the robustness of NPE for field-level weak lensing inference.

\begin{acknowledgments}
This material is based on work supported by the National Science Foundation under Grant No. 2209720 and the U.S. Department of Energy, Office of Science, Office of High Energy Physics under Award Number DE-SC0023714. The authors thank Xiangchong Li for helpful comments and technical assistance with AnaCal.

This paper has undergone internal review in the LSST Dark Energy Science Collaboration. The internal reviewers were Arun Kannawadi and Masaya Yamamoto.

The DESC acknowledges ongoing support from the Institut National de Physique Nucl\'eaire et de Physique des Particules in France; the Science \& Technology Facilities Council in the United Kingdom; and the Department of Energy and the LSST Discovery Alliance in the United States. DESC uses resources of the IN2P3 Computing Center (CC-IN2P3--Lyon/Villeurbanne - France) funded by the Centre National de la Recherche Scientifique; the National Energy Research Scientific Computing Center, a DOE Office of Science User Facility supported by the Office of Science of the U.S.\ Department of Energy under Contract No.\ DE-AC02-05CH11231; STFC DiRAC HPC Facilities, funded by UK BEIS National E-infrastructure capital grants; and the UK particle physics grid, supported by the GridPP Collaboration. This work was performed in part under DOE Contract DE-AC02-76SF00515.
\end{acknowledgments}

\begin{contribution}
TW and DT created the blissWL software package. TW, DT, CA, and JR devised the experiments. TW conducted the experiments and generated the figures. CA and JR supervised the project. TW, DT, CA, and JR wrote the text of this manuscript.
\end{contribution}

\section*{Data Availability}

Our fork of the \texttt{descwl-shear-sims} repository is publicly available at \url{https://github.com/timwhite0/descwl-shear-sims}. Code for simulating images for all five settings can be found in the \texttt{image\_generation} folder.

Our implementation of NPE is publicly available at \url{https://github.com/prob-ml/blissWL}. Code for reproducing the tables and figures in this paper can be found in the \texttt{images\_to\_maps/descwl} directory.


\bibliography{references}{}
\bibliographystyle{aasjournalv7}

\appendix
\crefalias{section}{appendix}
\setcounter{figure}{0}
\setcounter{table}{0}
\renewcommand{\thefigure}{\thesection\arabic{figure}}
\renewcommand{\thetable}{\thesection\arabic{table}}
\makeatletter
\@addtoreset{figure}{section}
\@addtoreset{table}{section}
\makeatother

\section{Neural network architecture} \label{appdx:networkarchitecture}

\Cref{fig:architecturediagram} illustrates how our neural network $f_\eta$ maps an image $x$ to variational parameters $\phi$. The network's inputs and outputs (outlined in black) are connected by a series of individual layers (shaded gray) and multi-layer residual blocks (shaded blue).

The backbone of the network consists of eight residual blocks. Each residual block transforms the network's intermediate representations based on an input channel count $c$, an output channel count $c^\prime$, and a stride. The resulting channel depth and spatial resolution of these intermediate representations are listed in italics on the left side of the diagram. While these tensors also have a batch dimension, we omit it from the diagram.

The panel on the right side of the diagram details the internal structure of a single residual block. Each block employs two 3x3 convolutional layers interspersed with group normalization and SiLU activation layers \citep{wu2018group}, as well as a skip connection composed of a 1x1 convolution and group normalization.

\vfill

\begin{minipage}{\textwidth}
    \centering
    \resizebox{0.9\textwidth}{!}{%
    \begin{tikzpicture}[
        inputoutput/.style={
        rectangle, ultra thick, draw=black, align=center, minimum width=1in
        },
        resblockoutline/.style={
        rectangle, ultra thick, dotted, rounded corners, draw=mediumteal, align=center, minimum width=2in
        },
        resblock/.style={
        rectangle, thick, rounded corners, draw=mediumteal, fill=mediumteal!10, align=center, minimum width=2in
        },
        preprocesslayer/.style={
        rectangle, thick, rounded corners, draw=pinkgray, fill=pinkgray!10, align=center, minimum width=1.75in
        },
        resblocklayer/.style={
        rectangle, thick, rounded corners, draw=pinkgray, fill=pinkgray!10, align=center, minimum width=1.75in
        },
        finallayer/.style={
        rectangle, thick, rounded corners, draw=pinkgray, fill=pinkgray!10, align=center, minimum width=1.5in
        },
        dimensions/.style={
        align=center, font=\itshape
        },
        arrow/.style={
        ->, thick, shorten <=0.1cm, shorten >=0.1cm
        },
    ]
        \node[resblock, minimum width = 3.75in] (resblocktitle) {ResidualBlock(c, c$^\prime$, s)};
        \node[inputoutput, below=0.15in of resblocktitle, xshift=-0.9in] (in) {Input\\
        c $\times$ h $\times$ w};
        \node[resblocklayer, below=0.15in of in] (conv1) {3x3 Conv\\
        c$_{\text{in}}$ = c, c$_{\text{out}}$ = c$^\prime$, stride = s};
        \node[resblocklayer, below=0.15in of conv1] (gn1) {GroupNorm};
        \node[resblocklayer, right=0.15in of gn1, yshift=-0.2in] (outerconv) {1x1 Conv\\
        c$_{\text{in}}$ = c, c$_{\text{out}}$ = c$^\prime$, stride = s};
        \node[resblocklayer, below=0.15in of outerconv] (outergn) {GroupNorm};
        \node[resblocklayer, below=0.15in of gn1] (silu1) {SiLU};
        \node[resblocklayer, below=0.15in of silu1] (conv2) {3x3 Conv\\
        c$_{\text{in}}$ = c$^\prime$, c$_{\text{out}}$ = c$^\prime$, stride = 1};
        \node[resblocklayer, below=0.15in of conv2] (gn2) {GroupNorm};
        \node[circle, thick, draw=gray, below=0.15in of gn2] (connection) {+};
        \node[resblocklayer, below=0.15in of connection] (silu2) {SiLU};
        \node[inputoutput, below=0.15in of silu2] (out) {Output\\
        c$^\prime$ $\times$ $\frac{\text{h}}{\text{s}}$ $\times$ $\frac{\text{w}}{\text{s}}$};
        \node[resblockoutline, fit={(resblocktitle) (in) (conv1) (gn1) (outerconv) (outergn) (silu1) (conv2) (gn2) (silu2) (connection) (out)}] (resblocktemplate) {};

        \draw[arrow, out=0, in=90] (in.east) to (outerconv.north);
        \draw[arrow] (outerconv.south) to (outergn.north);
        \draw[arrow, out=270, in=0] (outergn.south) to (connection.east);
        \draw[arrow] (in.south) to (conv1.north);
        \draw[arrow] (conv1.south) to (gn1.north);
        \draw[arrow] (gn1.south) to (silu1.north);
        \draw[arrow] (silu1.south) to (conv2.north);
        \draw[arrow] (conv2.south) to (gn2.north);
        \draw[arrow] (gn2.south) to (connection.north);
        \draw[arrow] (connection.south) to (silu2.north);
        \draw[arrow] (silu2.south) to (out.north);

        \node[resblock, left=0.25in of resblocktemplate] (rb3) {ResidualBlock(c=512, c$^\prime$=512, stride=2)};
        \node[resblock, above=0.25in of rb3] (rb2) {ResidualBlock(c=256, c$^\prime$=512, stride=2)};
        \node[resblock, above=0.25in of rb2] (rb1) {ResidualBlock(c=128, c$^\prime$=256, stride=2)};
        \node[resblock, above=0.25in of rb1] (rb0) {ResidualBlock(c=64, c$^\prime$=128, stride=2)};
        \node[resblock, below=0.25in of rb3] (rb4) {ResidualBlock(c=512, c$^\prime$=256, stride=2)};
        \node[resblock, below=0.25in of rb4] (rb5) {ResidualBlock(c=256, c$^\prime$=128, stride=2)};
        \node[resblock, below=0.25in of rb5] (rb6) {ResidualBlock(c=128, c$^\prime$=64, stride=2)};
        \node[resblock, below=0.25in of rb6] (rb7) {ResidualBlock(c=64, c$^\prime$=32, stride=2)};

        \draw[thick, teal, dotted] (rb0.east) to (resblocktemplate.north west);
        \draw[thick, teal, dotted] (rb0.east) to (resblocktemplate.south west);
        
        \draw[arrow] (rb0.south) to node[midway, left=1.25in, dimensions] {128 $\times$ 1024 $\times$ 1024} (rb1.north);
        \draw[arrow] (rb1.south) to node[midway, left=1.25in, dimensions] {256 $\times$ 512 $\times$ 512} (rb2.north);
        \draw[arrow] (rb2.south) to node[midway, left=1.25in, dimensions] {512 $\times$ 256 $\times$ 256} (rb3.north);
        \draw[arrow] (rb3.south) to node[midway, left=1.25in, dimensions] {512 $\times$ 128 $\times$ 128} (rb4.north);
        \draw[arrow] (rb4.south) to node[midway, left=1.25in, dimensions] {256 $\times$ 64 $\times$ 64} (rb5.north);
        \draw[arrow] (rb5.south) to node[midway, left=1.25in, dimensions] {128 $\times$ 32 $\times$ 32} (rb6.north);
        \draw[arrow] (rb6.south) to node[midway, left=1.25in, dimensions] {64 $\times$ 16 $\times$ 16} (rb7.north);
        
        \node[preprocesslayer, above=0.25in of rb0] (preprocess_silu) {SiLU};
        \node[preprocesslayer, above=0.25in of preprocess_silu] (preprocess_gn) {GroupNorm};
        \node[preprocesslayer, above=0.25in of preprocess_gn] (preprocess_conv) {5x5 Conv\\
        c$_{\text{in}}$ = 3, c$_{\text{out}}$ = 64, stride = 1};
        \node[inputoutput, above=0.25in of preprocess_conv, align=center] (images) {Image $x$};
        
        \draw[arrow] (images.south) to node[midway, left=1.25in, dimensions] {3 $\times$ 2048 $\times$ 2048} (preprocess_conv.north);
        \draw[arrow] (preprocess_conv.south) to node[midway, left=1.25in, dimensions] {64 $\times$ 2048 $\times$ 2048} (preprocess_gn.north);
        \draw[arrow] (preprocess_gn.south) to node[midway, left=1.25in, dimensions] {64 $\times$ 2048 $\times$ 2048} (preprocess_silu.north);
        \draw[arrow] (preprocess_silu.south) to node[midway, left=1.25in, dimensions] {64 $\times$ 2048 $\times$ 2048} (rb0.north);

        \node[finallayer, below=0.25in of rb7] (final) {Fully connected\\
        c$_{\text{in}}$ = 32$\cdot$8$\cdot$8, c$_{\text{out}}$ = 4};
        \node[inputoutput, below=0.25in of final] (varparams) {Variational parameters $\phi$};
        
        \draw[arrow] (rb7.south) to node[midway, left=1.25in, dimensions] {32 $\times$ 8 $\times$ 8} node[midway, right=0.025in] {flatten} (final.north);
        \draw[arrow] (final.south) to node[midway, left=1.25in, dimensions] {4} (varparams.north);
    \end{tikzpicture}
    }%
    \captionof{figure}{The architecture of our neural network $f_\eta$.}
    \label{fig:architecturediagram}
\end{minipage}

\newpage

\section{Inference timing comparison} \label{appdx:timingcomparison}

\begin{minipage}{\textwidth}
\centering
    \begin{tabular}{lcc}
        \toprule
        & \textbf{AnaCal} & \textbf{NPE} \\
        \midrule
        Setting 1 & 0.215 & 0.176 \\
        Setting 2 & 2.734 & 0.179 \\
        Setting 3 & --- & 0.181 \\
        Setting 4 & --- & 0.184 \\
        Setting 5 & --- & 0.186 \\
        \bottomrule
    \end{tabular}%
    \captionof{table}{Average number of seconds required to produce point estimates (AnaCal) or variational distribution parameters (NPE) for a single image, computed on a subset of 56 images for each setting. For this comparison, NPE inferences were computed using the pretrained neural network weights on one NVIDIA GeForce RTX 2080 Ti GPU; AnaCal was run with 28 parallel workers on a 32-core CPU.}
    \label{table:timing}
\end{minipage}

\newpage

\section{Posterior density plots} \label{appdx:densityplots}

\begin{minipage}{\textwidth}
    \centering
    \begin{tikzpicture}
    \node (s1) {\includegraphics[width=\textwidth]{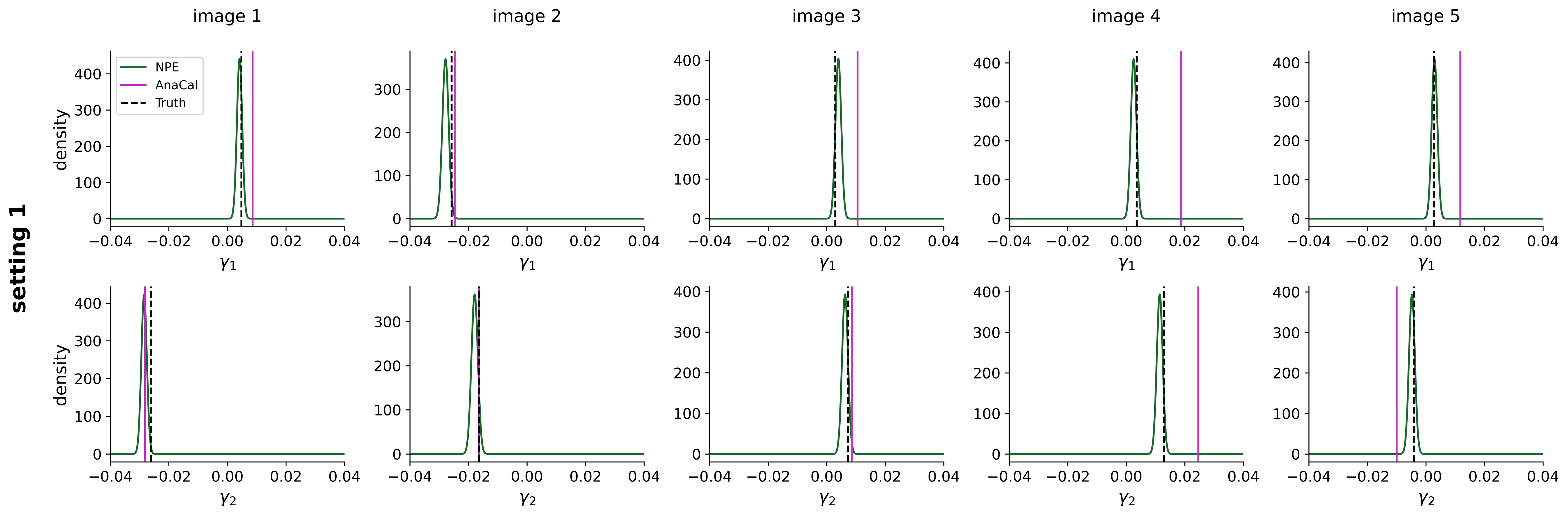}};
    \node[below=of s1] (s2) {\includegraphics[width=\textwidth]{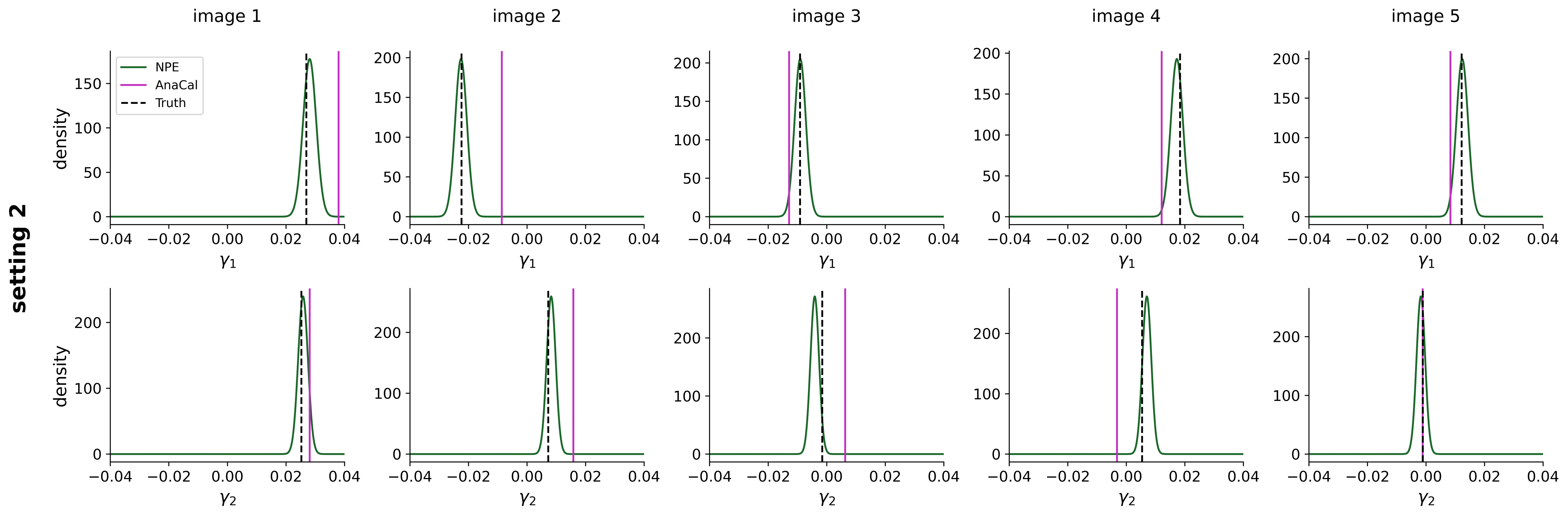}};
    \end{tikzpicture}
    \captionof{figure}{Inferred shear for five randomly selected images from settings 1 and 2. The posterior densities inferred by NPE are represented as solid green curves. The AnaCal point estimates are represented as solid pink vertical lines. The true shears are represented as dashed black vertical lines.}
    \label{fig:densityplots12}
\end{minipage}

\begin{minipage}{\textwidth}
    \centering
    \begin{tikzpicture}
    \node (s3) {\includegraphics[width=\textwidth]{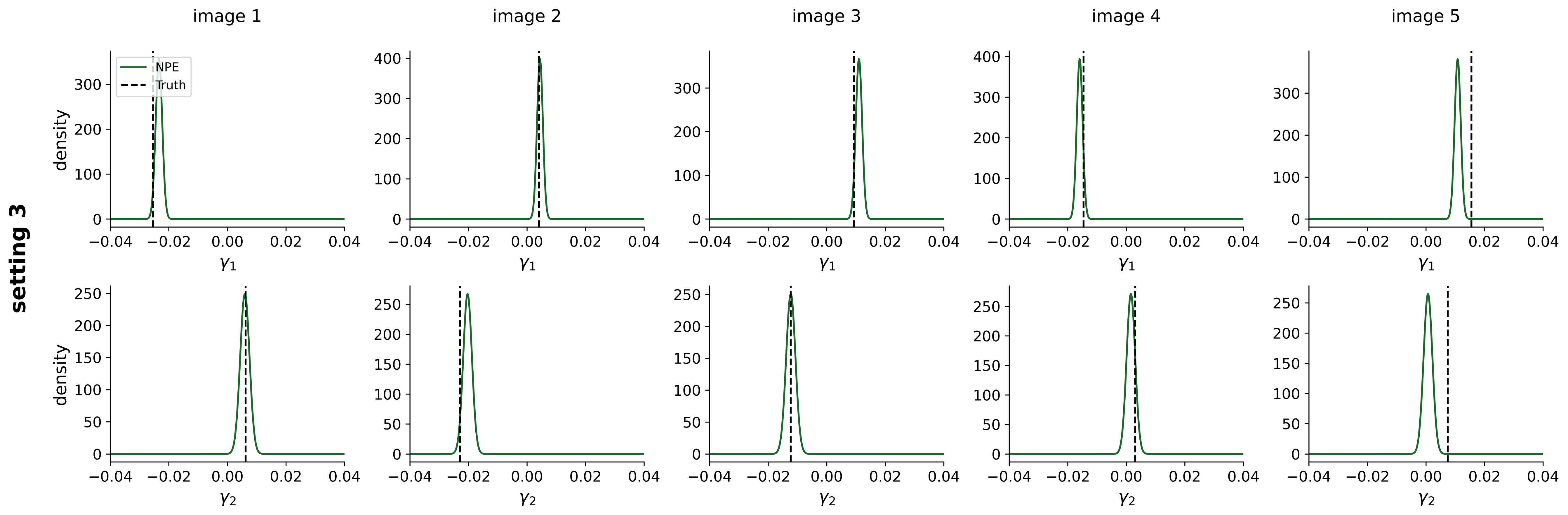}};
    \node[below=of s3] (s4) {\includegraphics[width=\textwidth]{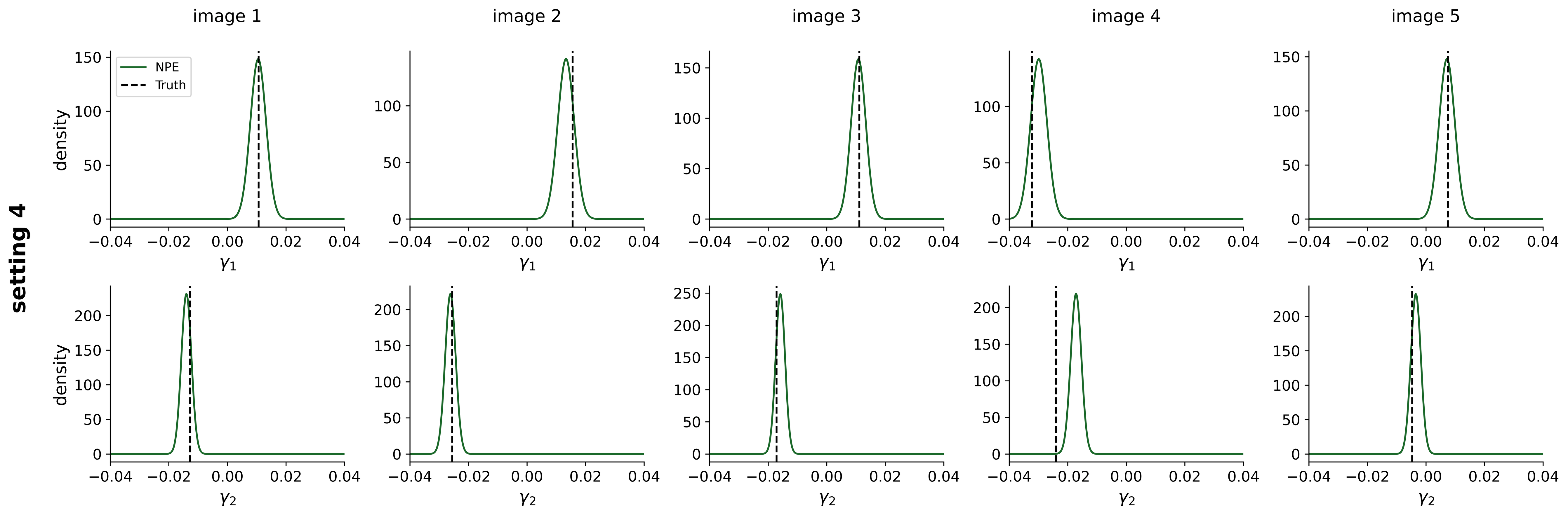}};
    \node[below=of s4] (s5) {\includegraphics[width=\textwidth]{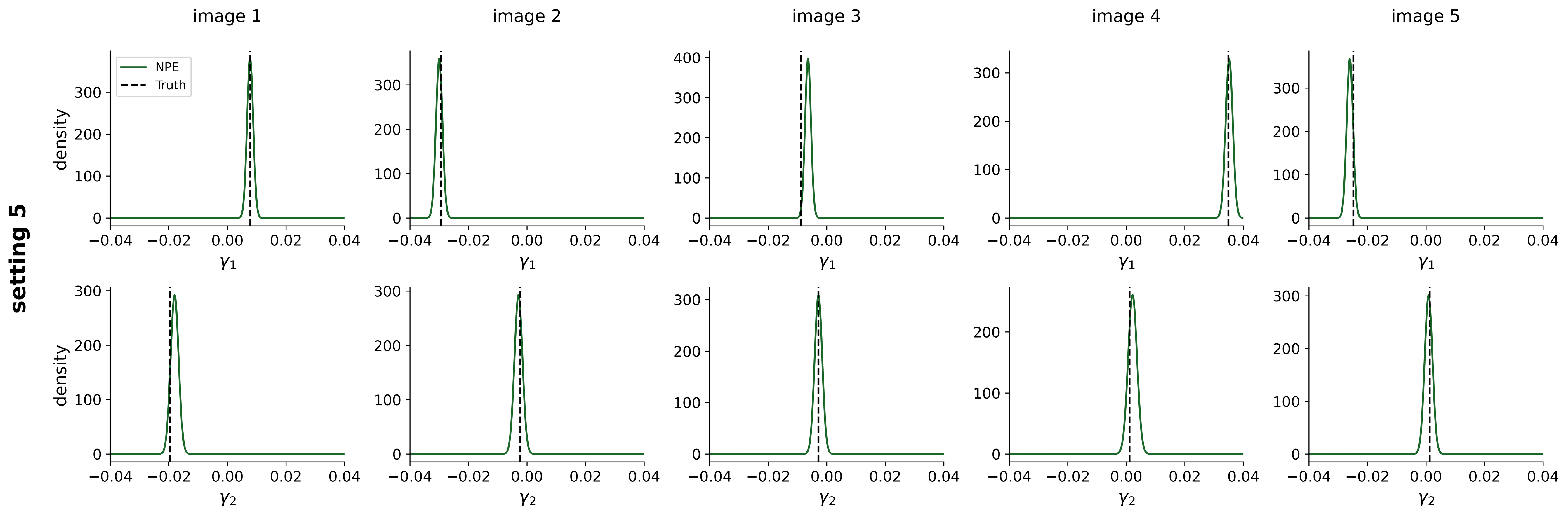}};
    \end{tikzpicture}
    \captionof{figure}{Inferred shear for five randomly selected images from settings 3, 4, and 5. The posterior densities inferred by NPE are represented as solid green curves. The true shears are represented as dashed black vertical lines.}
    \label{fig:densityplots345}
\end{minipage}

\end{document}